%% file: main.tex
\documentclass[12pt,3p,doublespacing]{elsarticle}
\journal{Journal of the Taiwan Institute of Chemical Engineers}
\usepackage{times}
\emergencystretch=\maxdimen
\usepackage{sectsty}
\sectionfont{\color{blue}\large\MakeUppercase}
\subsectionfont{\color{blue}}
\biboptions{sort&compress}
\usepackage{hyperref}
 \hypersetup{
 	plainpages = false, 
	bookmarksopen = true,
	bookmarksnumbered = true,
	breaklinks = true,
	linktocpage,
	colorlinks = true,
	linkcolor = purple,
	urlcolor  = black,
	citecolor = purple,
	}
\usepackage{multicol}
\usepackage{multirow}
\usepackage{imakeidx}
\makeindex
\usepackage{nomencl}
\makenomenclature
\usepackage[xindy]{glossaries}
\makeglossaries
\usepackage{graphics}
\usepackage{graphicx}
\usepackage{epsf,psfrag}
\usepackage{epstopdf}
\usepackage[para]{threeparttable}
\usepackage{subfigure}
\usepackage{epsfig}
\usepackage{pdflscape}
\usepackage{rotating}
\usepackage{lscape}
\usepackage{float}
\usepackage{stfloats}
\usepackage{textgreek}
\usepackage{longtable}
\usepackage{lscape}
\usepackage{capt-of}
\usepackage{comment}
\usepackage{tablefootnote}
\usepackage{footnote}
\usepackage{caption}
\usepackage[autopunct=true]{csquotes}
\usepackage{rotating}
\usepackage{txfonts}
\usepackage[normalem]{ulem}
\usepackage{resizegather}
\usepackage{siunitx}
\usepackage{setspace}
\usepackage{color,soul}
\usepackage[usenames,dvipsnames]{xcolor}
\usepackage{amsfonts,amsmath,amssymb}
\usepackage{latexsym,array}
\usepackage{textcomp}

\newcommand{\RomanNumeralCaps}[1]
\linenumbers
\makesavenoteenv{tabular}
\makesavenoteenv{table}
\usepackage[left,displaymath, mathlines]{lineno}
\DeclareMathAlphabet{\mathpzc}{OT1}{pzc}{m}{it}
\def\fig{Figure~}
\def\figs{Figures~}
\def\eqn{Eq.~}
\def\eqns{Eqs.~}
\def\tab{Table~}

\newcommand\mm[1]{$#1$}  
\newcommand\subt{_{\text{t}}}
\newcommand\subb{_{\text{b}}}
\newcommand\subr{_{\text{r}}}
\newcommand\sur{S_{\text{r}}}
\newcommand\sut{S_{\text{t}}}
\newcommand\sub{S_{\text{b}}}
\newcommand\smin{_{\text{min}}}
\newcommand\smax{_{\text{max}}}
\newcommand\avg{_{\text{avg}}}

\newcommand{\myvec}[1]{\mathbf{#1}}     
\def\tsc#1{\csdef{#1}{\textsc{\lowercase{#1}}\xspace}}
\tsc{WGM}
\tsc{QE}
\tsc{EP}
\tsc{PMS}
\tsc{BEC}
\tsc{DE}
\newcommand{\del}[1]{{\textcolor{orange}{\expandafter\sout\expandafter{#1}}}}

\newcommand{\rev}[1]{\textcolor{black}{#1}}     
\newcommand{\htxt}[1]{\textcolor{black}{\uline{#1}}}     
\newcommand{\ttxt}[1]{\textbf{\color{blue}\large\MakeUppercase{#1}}}     

\graphicspath{{../}{Figures/}{Figures/Validation/}}
\DeclareGraphicsExtensions{.png, .PNG,.pdf,.PDF,.jpg,.JPG,.eps,.EPS}
\begin{document}

%
%
\setcounter{page}{1}
\begin{frontmatter} 
%
%
%
%
%
%
\title{\ttxt{Electroviscous effects in pressure-driven flow of electrolyte liquid through an asymmetrically charged non-uniform microfluidic device}}
%
%
%
\author[labela]{Jitendra {Dhakar}}
\author[labela]{Ram Prakash {Bharti}\corref{coradd}}\ead{rpbharti@iitr.ac.in}
\address[labela]{Complex Fluid Dynamics and Microfluidics (CFDM) Lab, Department of Chemical Engineering,\\ Indian Institute of Technology Roorkee, Roorkee - 247667, Uttarakhand, INDIA}
%
%
\cortext[coradd]{\textit{Corresponding author. }}
%
\begin{abstract}
\fontsize{11}{13pt}\selectfont
Micro-scale systems depict a different flow behavior from the macro-scale systems due to more vital surface forces such as surface tension, electrical charges, magnetic field, etc., which significantly affect the micro-scale flow. Further, among others, electrokinetic phenomena play a significant role at the micro-scale for controlling practical microfluidic applications. Therefore, it is essential to understand the fluid dynamics in micron-sized channels to develop efficient and reliable microfluidic devices. 
The electroviscous effects in pressure-driven flow of electrolyte liquid through an asymmetrically charged contraction-expansion (4:1:4) slit microfluidic device have been investigated numerically. The mathematical model (i.e., Poisson's, Navier-Stokes, and Nernst-Planck equations) is solved using the finite element method to obtain the electrical potential, velocity, pressure, ion concentration fields, excess charge, an induced electric field strength for the following ranges of parameters: Reynolds number (\mm{Re=0.01}), Schmidt number (\mm{\mathit{Sc}=1000}), inverse Debye length (\mm{2\le K\le 20}), top wall surface charge density (\mm{4\le \sut\le 16}), surface charge ratio (\mm{0\le \sur\le 2}) and contraction ratio (\mm{d_{\text{c}}=0.25}). 
Results show that the charge asymmetry at the different walls of the microfluidic device plays a significant role on the induced electric field development and microfluidic hydrodynamics. The total potential (\mm{|\Delta U|}) and pressure drop (\mm{|\Delta P|}) maximally increase by 197.45\% and 25.46\%, respectively with asymmetry of the charge. The electroviscous correction factor (ratio of apparent to physical viscosity) maximally changes by 20.85\% (at \mm{K=2}, \mm{\sut=16} for \mm{0\le \sur\le 2}), 34.16\% (at \mm{\sut=16}, \mm{\sur=2} for \mm{2\le K\le 20}), and 39.13\% (at \mm{K=2}, \mm{\sur=2} for \mm{0\le\sut\le16}). Thus, charge asymmetry ($0\le S_\text{r}\le 2$) remarkably influences the fluid flow in the microfluidic devices, which is used for controlling the microfluidic processes, such as, mixing efficiency, heat, and mass transfer rates. Further, a simpler analytical model is developed to predict the pressure drop in electroviscous flow considering asymmetrically charged surface, based on the Poiseuille flow in the individual uniform sections and pressure losses due to orifice, estimates the pressure drop 1--2\% within the numerical results. The robustness of this model enables the use of present numerical results for design aspects in the microfluidic applications.
\end{abstract}
\end{frontmatter}
%
\section{Introduction}
\label{sec:intro}
%
\noindent 
The rapid development of the fabrication techniques of micro-electro-mechanical systems has enhanced the use and popularity of microfluidic devices in various biomedical and engineering applications such as micro-heat pumps, micro-heat sinks, DNA analysis, drug screening, drug delivery systems, Lab-on-a-chip, bio-analysis, cell cultivation and droplet generation \citep{li2008encyclopedia,bhushan2007springer,lin2011microfluidics,
foudeh2012microfluidic,vladisavljevic2013industrial,nguyen2013design,Tehranirokh2013,bruijns2016microfluidic,damiati2018microfluidic,he2019preparation,ortseifen2020microfluidics,kuo2020biomaterial,Berlanda2021,kim2021thin,li2021microfluidic,Venkat2021,lan2022application,adam2023integration,Venkat2023}. 
%
Micro-scale systems depict a different flow behavior from the macro-scale systems due to more vital surface forces such as surface tension, electrical charges, magnetic field, etc., which significantly affect the micro-scale flow. Therefore, it is essential to understand the fluid dynamics in micron-sized channels to develop efficient and reliable microfluidic devices.

\noindent
{Most solid surfaces (PDMS, glass, other materials) contain electrostatic charge, i.e., surface electrical potential (\mm{\zeta_\text{s}}). The electrokinetic phenomena arise when these charged surfaces interact with liquid electrolyte solutions. The charged solid surfaces attract counter-ions and repeal co-ions; thus, the rearrangement of the ions near the surface forms an `electrical double layer' (EDL) \citep{hunter2013zeta,li2001electro,Masliyah06,Tadros2013}. It consists of the compact (or Stern) and diffuse layer; the interface located between these layers of the EDL is defined as the shear plane. 
Zeta potential (\mm{\zeta}) is the potential at the shear plane (or slipping plane), and it continuously decays in the diffuse layer from the shear plane to the bulk liquid \citep{Hsu2016,dhakar2022electroviscous}. The convective flow of mobile ions in the diffuse layer of EDL due to applied pressure-driven flow (PDF) results in a `streaming current'. This flow induces a `streaming potential' that imposes a back-flow of counter ions in the EDL opposite to PDF and generates a `conduction current'. These ions also drive liquid with them and retards the primary PDF in the microfluidic device. Thus, the liquid flow shows higher viscosity than conventional PDF at a fixed volumetric flow rate; this effect is known as the `electroviscous effect' (EVE) \citep{hunter2013zeta,atten1982electroviscous}.}

\noindent
Earlier experimental and numerical studies have explored electroviscous effects in symmetrically/uniformly charged  uniform microfluidic devices such as parallel-plate \citep{burgreen1964electrokinetic,mala1997flow,mala1997heat,chun2003electrokinetic,ren2004electroviscous,chen2004developing,joly2006liquid,wang2010flow,jamaati2010pressure,zhao2011competition,tan2014combined,jing2015electroviscous,matin2016electrokinetic,jing2017non,matin2017electroviscous,kim2018analysis,mo2019electroviscous,li2021combined,Li2022}, rectangular \citep{yang1998modeling,ren2001electro,li2001electro}, cylindrical \citep{rice1965electrokinetic,levine1975theory,bowen1995electroviscous,brutin2005modeling,bharti2009electroviscous,jing2016electroviscous},  and elliptical \citep{hsu2002electrokinetic}. 
%
%
Researchers have also explored the electroviscous effects in pressure-driven no-slip flow of ionic/electrolyte liquids through symmetrically charged non-uniform (i.e., contraction-expansion) microfluidic devices such as parallel-plate/slit \citep{davidson2007electroviscous,Berry2011}, rectangular \citep{davidson2008electroviscous}, and cylindrical \citep{bharti2008steady,davidson2010electroviscous}. 
%
\citet{Berry2011} have analyzed the effect of wall permittivity with uniform charge density on liquid-solid interface on the electroviscous flow, under otherwise identical conditions \citep{davidson2007electroviscous}.
Further, \citet{dhakar2022electroviscous,dhakar2022slip} have investigated electroviscous effects in the charge-dependent slip flow of an electrolyte solution through symmetrically charged non-uniform (i.e., contraction-expansion) slit microchannel device.  
Combined altogether, these studies  \citep{davidson2007electroviscous,davidson2008electroviscous,bharti2008steady,bharti2009electroviscous,davidson2010electroviscous,Berry2011,dhakar2022electroviscous,dhakar2022slip} have shown the stronger effects of inverse Debye length ($2\le K\le 20$), surface charge density ($4\le S\le 16$) and slip length ($0\le B_\text{0}\le 0.20$) for a fixed volumetric flow rate ($Q$) on the governing fields, i.e., total electrical potential ($U$), velocity ($\mathbf{V}$), ion-concentrations ($n_{\pm}$), pressure ($P$), excess charge ($n^\ast$), and induced electric field strength ($E_\text{x}$) in symmetrically charged microfluidic device.
These studies have observed substantial electroviscous effects in non-uniform compared to uniform geometries under otherwise identical ranges of conditions. Further, slip enhanced the electroviscous effects compared to the no-slip case, irrespective of the microfluidic geometry.
Simple predictive models to calculate the pressure drop have been presented for no-slip \citep{davidson2007electroviscous,bharti2008steady,davidson2008electroviscous,davidson2010electroviscous} and charge-dependent slip \citep{dhakar2022electroviscous,dhakar2022slip}  flow through symmetrically charged non-uniform microfluidic devices by accounting for the pressure drop in the Poiseuille flow through the uniform channel, and that in creeping flow through thin orifice \citep{Sisavath2002}. The model predicted pressure drop (and electroviscous correction factor) within $\pm$ 5\% and $\pm$ 2 -- 4\% of their numerical results for no-slip and charge-dependent slip flow.

\noindent 
Furthermore, few studies have explored electroviscous effects (EVEs) in asymmetrically or heterogeneously charged uniform microfluidic devices. 
For instance, \citet{xuan2008streaming} has theoretically investigated the streaming potential and EVEs in a heterogeneously charged uniform microchannel with two types of surface patterns  (i.e., $q\parallel\nabla P$ and $q\bot\nabla P$). 
\citet{sailaja2019electroviscous} have shown significant effects of asymmetric wall zeta potential, i.e., channel walls made of different materials, on the hydrodynamics of power-law fluid flow through uniform slit microchannels. They have concluded that the asymmetry of zeta potentials has strongly affected the streaming potential in microfluidic devices. 

\noindent 
To the best of our knowledge, the effect of charge asymmetry between the upper and lower walls on electroviscous flow through non-uniform microfluidic geometries are unexplored in the literature.
Therefore, this study has investigated the electrolyte liquid flow through asymmetrically charged non-uniform (i.e., contraction-expansion) slit microfluidic device. A finite element method (FEM) is used to numerically solve the mathematical equations to obtain the flow, electrical potential, and ion concentration fields. The detailed results are presented and discussed in-depth in terms of the total electrical potential ($U$), excess charge ($n^\ast$), induced electric field strength ($E_\text{x}$), and pressure ($P$) distributions in the microfluidic device for the broader ranges of the flow governing parameters ($2\le K\le 20$, $4\le \sut\le 16$ and $0\le \sur\le 2$).
A simple pseudo-analytical model has been presented to predict the pressure drop (and electroviscous correction factor). This model can be used to design efficient and reliable microfluidic devices/chips for various engineering and biomedical applications.
%
%
\begin{figure}[!b]
	\centering
	\subfigure[]{
	\includegraphics[width=1\linewidth]{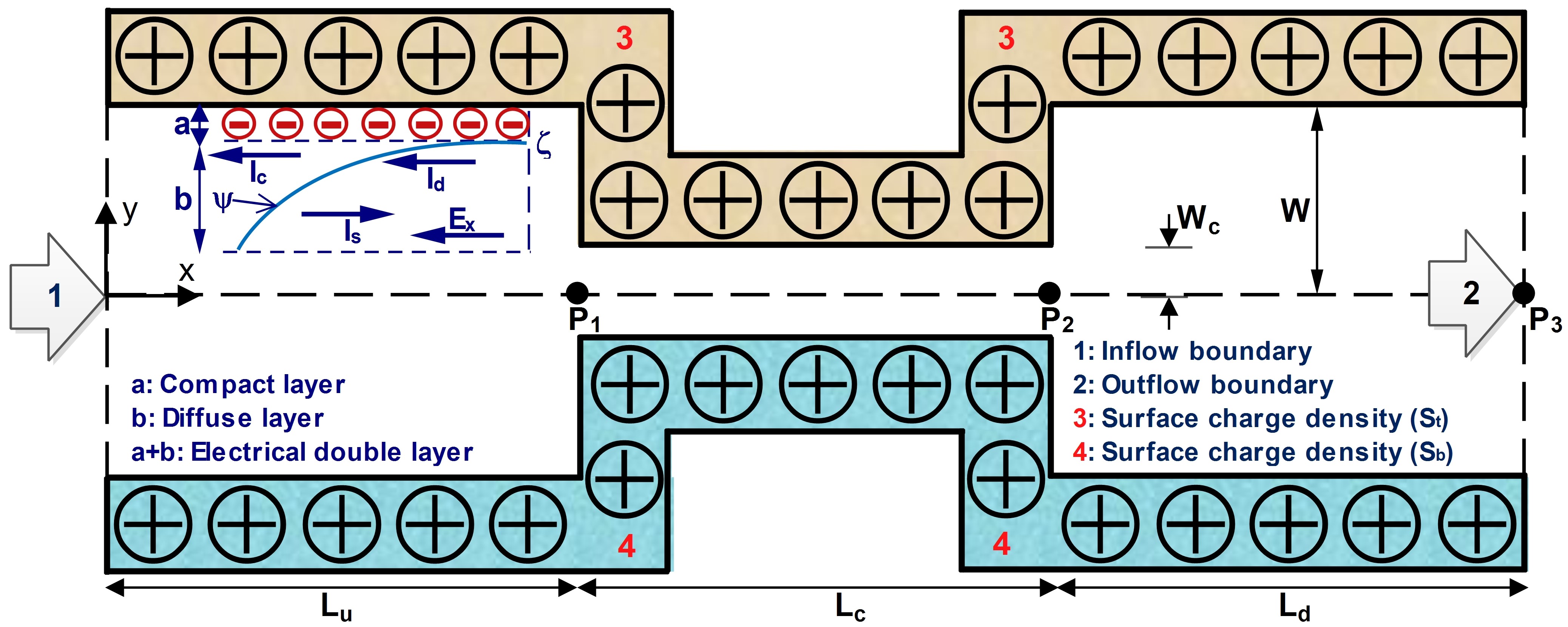}\label{fig:1a}}
	\subfigure[]{\includegraphics[width=0.8\linewidth]{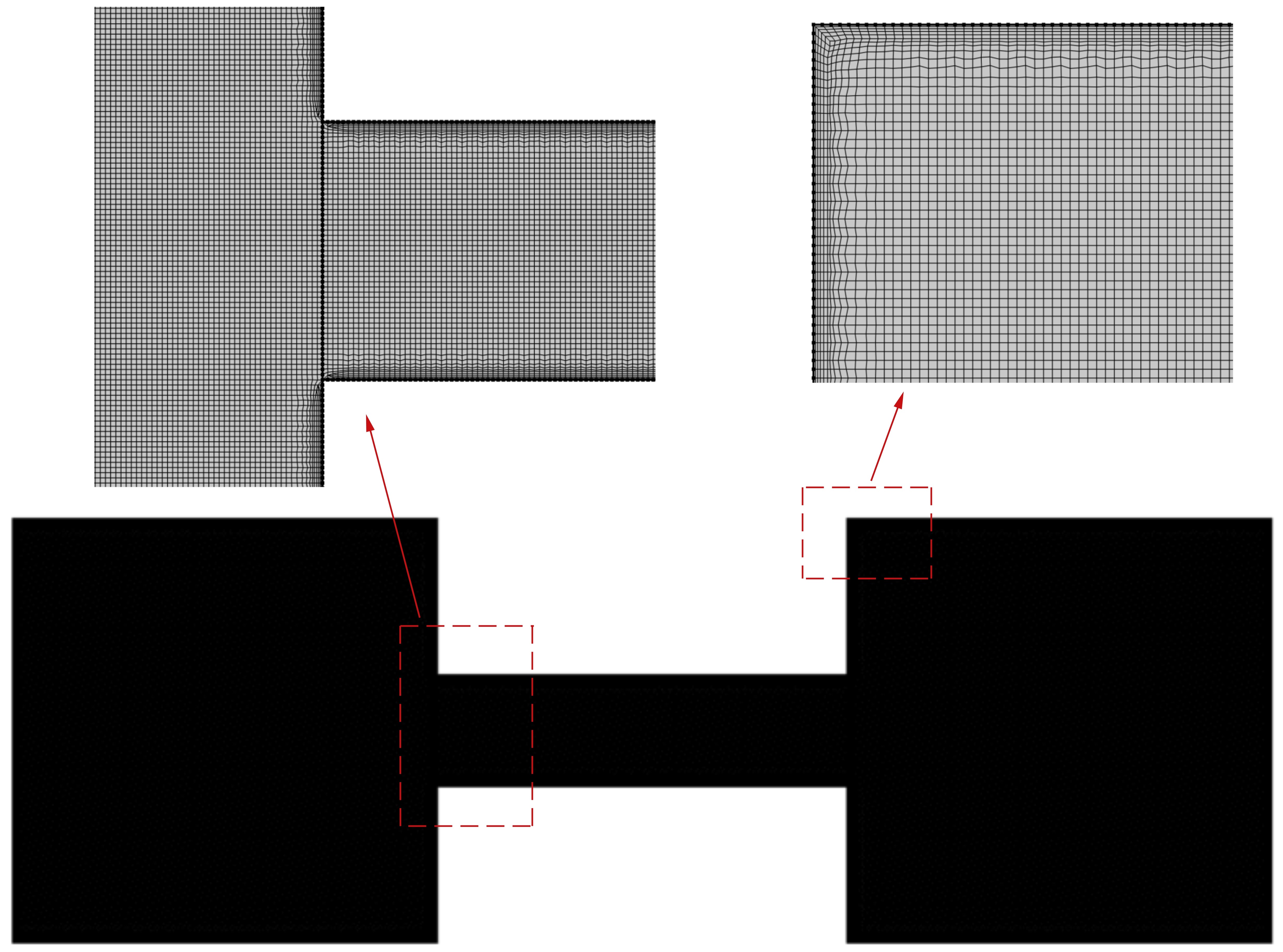}\label{fig:1b}}
	\caption{\rev{Representation of (a) electroviscous flow (EVF) through asymmetrically charged contraction-expansion microfluidic device (b) mesh distribution in the computational domain}.}
	\label{fig:1}
\end{figure}
\section{Physical and mathematical modelling}
%
\noindent 
Consider the two-dimensional steady laminar fully developed flow of electrolyte liquid through a asymmetrically charged non-uniform (i.e., contraction-expansion) slit microfluidic device, as shown in \fig\ref{fig:1a}. The contraction section ($L_{\text{c}} \times 2W_{\text{c}}$, $\mu$m$^2$) is situated in between the upstream ($L_{\text{u}} \times 2W$, $\mu$m$^2$) and downstream ($L_{\text{d}} \times 2W$, $\mu$m$^2$) sections. 
The total length of the microchannel is $L~(=L_{\text{u}}+L_{\text{c}}+L_{\text{d}})$ $\mu$m, and the contraction ratio is defined as $d_\text{c}=(W_\text{c}/W)$. 
\noindent
The liquid is assumed to have symmetric ($1{:}1$) anions and cations with equal ionic valence ($z_{+}=-z_{-}=z$) and diffusivity ($\mathcal{D}_{+}=\mathcal{D}_{-}=\mathcal{D}$, m$^2$/s). The geometric mean ionic concentration is $n_{\text{0}}$, mol/L  \citep{harvie2012microfluidic,davidson2016numerical,dhakar2022electroviscous}, and the average inlet flow velocity of electrolyte liquid is $\overline{V}$, m/s.
The top and bottom walls of the microchannel are assumed to have a uniform but unequal surface charge density as $\sigma\subt$, C/m$^2$ and $\sigma\subb$, C/m$^2$, respectively. The liquid is considered Newtonian and incompressible, i.e., physical properties of the liquid such as density ($\rho$, kg/m$^2$), viscosity ($\mu$, Pa.s), and dielectric constant ($\varepsilon\subr$) are spatially uniform. The dielectric constant of the device material is assumed to be very small comparatively to the liquid  ($\varepsilon_{\text{r,w}} \lll \varepsilon\subr$). 
\subsection{Governing equations}
\noindent 
The electroviscous (EV) flow can be modeled by the conservations of electrical potential, the mass of charged ionic species, mass, and momentum expressed by the Poisson's, Nernst-Planck (NP), and Navier-Stokes (NS) equations. \rev{The detailed mathematical model, in both dimensional and dimensionless forms, is described elsewhere \citep[refer \eqns 1 to 4, and A.1 to A.8 in][]{dhakar2022electroviscous}}. 
\\\noindent 
The electrohydrodynamic theory suggest that the total electrical potential ($U$) in the electrokinetic flow can be expressed as sum of two kinds of potential, (i) EDL potential ($\psi$) and (ii) streaming potential  ($\phi=-xE_x$), as follows  \citep{davidson2007electroviscous,bharti2008steady,bharti2009electroviscous,dhakar2022electroviscous}
\begin{gather}
	U(x,y) = \psi - x E_{\text{x}}
	\label{eq:2a}
\end{gather}
where $\psi$, $E_{\text{x}}$  and $(x E_{\text{x}})$ are the EDL potential (V), induced electric field strength (V/m) in axial direction, and streaming potential (V), respectively. Since EDL potential ($\psi$) varies in the direction normal to the charged surface, and the streaming potential ($\phi$) varies axially in the direction of the convective flow, the two potentials ($\psi$ and $\phi$) can therefore be split \citep{davidson2007electroviscous,bharti2008steady,bharti2009electroviscous,dhakar2022electroviscous} in the case of the uniform microfluidic geometry as they are independent of each other. The Poisson (or Poisson-Boltzmann) equation can thus be written in terms of EDL potential ($\psi$) and external body force term in the Navier-Stokes (or Cauchy) momentum equation in terms of streaming potential  ($\phi$) \citep{bharti2008steady}. On the other hand, in the case of non-uniform (i.e., contraction-expansion)  geometries, such as the present study, splitting the two potentials ($\psi$ and $\phi$) is impossible. Thus, Poisson-Boltzmann and Navier-Stokes equations are expressed using total electrical potential  ($U$).
\noindent
The governing model equations have been rendered dimensionless by using the following scaling variables: $W$, $\overline{V}$,  ($W/\overline{V}$), $\rho\overline{V}^2$, $U_{\text{c}}~(=k_{\text{B}}T/ze)$, and $n_{\text{0}}$ for length, velocity, time, pressure, electrical potential, and number density of ions, respectively. 
The scaling analysis results in the dimensionless groups as follows.
\begin{gather}
	Re=\frac{\rho\overline{V}W}{\mu}, \qquad
	\mathit{Sc}=\frac{\mu}{\rho \mathcal{D}}, \qquad
	Pe =Re\times\mathit{Sc}, \qquad
	\beta=\frac{\rho \varepsilon_{\text{0}}\varepsilon\subr U_{\text{c}}^2}{2\mu^2}, \qquad 
	K^2=\frac{2W^2zen_{\text{0}}}{\varepsilon_{\text{0}}\varepsilon\subr U_{\text{c}}}
	\label{eq:1}
\end{gather}
where $Re$, $\mathit{Sc}$, $Pe$, $\beta$ and $K$ are the Reynolds, Schmidt and Peclet numbers, liquid parameter and inverse Debye length ($\lambda_{\text{D}}^{-1}$), respectively. 
Here, $k_{\text{B}}$, $\varepsilon_{\text{0}}$, $e$, and $T$ are the  Boltzmann constant, permittivity of free space, elementary charge of a proton, and temperature, respectively.

\noindent
Subsequently, the dimensionless form of the governing model is written as follows.
\noindent 
The electrical potential distribution can be described by Poisson’s equation that relates the total electrical potential ($U$) and  charge density of ions ($\rho_{\text{e}}$) as follows.
\begin{gather}
	\nabla^2U=-\frac{1}{2}K^2n^{\ast}\qquad\text{where}\qquad 
	n^{\ast} = (n_{+}-n_{-})
	\label{eq:2}
\end{gather}
where $n^{\ast}$ is the excess charge (or excess ionic density) which is equal to the dimensionless $\rho_{\text{e}}$ \citep[$\rho_{\text{e}} = \sum_{j=1}^N\rho_{\text{e,j}}$ where $\rho_{\text{e,j}}=z_jen_j$, refer \eqns (A.2) in][]{dhakar2022electroviscous} for the symmetric electrolyte, and $n_{\text{j}}$ is the local number density of $\text{j}^{\text{th}}$ ionic species. 

\noindent
The ionic species ($n_{\text{j}}$) conservation can be described by the Nernst-Planck  equation (NPE) as follows.
\begin{gather}
    \left[\frac{\partial n_{\text{j}}}{\partial t}+\nabla\cdot(\myvec{V}n_{\text{j}})\right]={Pe^{-1}}\left[\nabla^2n_{\text{j}}\pm\nabla\cdot(n_{\text{j}}\nabla U)\right]
	\label{eq:3}
\end{gather}
where $\myvec{V}$ is the velocity vector field. 

\noindent
The conservation of the flow field (i.e., velocity and pressure fields) is expressed by Cauchy momentum (with an extra electrical body force term) and mass continuity equations as follows.
\begin{gather}
	\left[\frac{\partial \mathbf{V}}{\partial t}+\nabla\cdot(\myvec{V}\myvec{V})\right] = 
	- \nabla P+{Re^{-1}}\nabla \cdot\left[\nabla\myvec{V}
	+ (\nabla\myvec{V})^T\right]
	-	\underbrace{\beta Re^{-2}{K}^2n^{\ast}\nabla U}_{\myvec{F}_{\text{e}}}
	\label{eq:4}
	\\
	\nabla\cdot\myvec{V}=0 \label{eq:5}
\end{gather}
where $\myvec{F_\text{e}}$ and $P$ are the electrical body force and pressure, respectively.
The coupled field equations (\eqns\ref{eq:2} to \ref{eq:5}) are subjected with the following physical boundary conditions.
\subsection{Boundary conditions}
\noindent 
%
%
The boundary conditions (BC) in dimensionless form \rev{\citep[refer \eqns A.9 to A.16 in][for dimensional form]{dhakar2022electroviscous}} for the present problem are described as follows.

\noindent
[BC-1]: 
The fully developed velocity and ionic species concentration fields at the inflow ($x=0$) boundary are imposed  as follows.
%
\begin{gather}
	V_{\text{x}}=V_{\text{0}}(y), \qquad
	V_{\text{y}}=0, \qquad
	n_{\pm}=\exp\left[\mp\psi(y)\right]
	\label{eq:6} 
\end{gather}
where $V_{\text{0}}$ and $\psi$ are the fully developed velocity and EDL potential fields that are obtained by the numerical solution for the steady, fully developed electroviscous flow of symmetric electrolytes through the two-dimensional (2-D)  uniform slit \citep{dhakar2022electroviscous,davidson2007electroviscous}. 
\\\noindent 
The fully developed inlet condition in this work has been applied to shorten the computational domain and reduce the computational efforts without any loss of generality of the physics, as a uniform in-flow condition requires an additional length of computational domain to achieve the fully developed flow field. The computations over the extra length (i.e., flow domain) and achieving fully developed flow require significantly more computational efforts.
%

\noindent
[BC-2]:  
The axial gradients of velocity and ionic species concentration fields are zero  (\eqn\ref{eq:8}) at the outflow ($x=L$) boundary open to ambient. 
\begin{gather}
	\frac{\partial \myvec{V}}{\partial \myvec{n}\subb} = 0,
	\qquad
	\frac{\partial n_{\text{j}}}{\partial \myvec{n}\subb} = 0,
	\qquad
	P =0
	\label{eq:8}
\end{gather}
where $\myvec{n}\subb$ is the unit vector outward normal to the boundary.

\noindent
[BC-3]: The `current continuity' condition (\eqn\ref{eq:7}), i.e., the zero net current ($I_\text{net} = \nabla\cdot I = 0$) in the axial direction at the steady-state, is satisfied at both the inflow ($x=0$) and outflow ($x=L$) boundaries to impose the uniform axial potential gradient \citep{davidson2007electroviscous,bharti2008steady,bharti2009electroviscous,dhakar2022electroviscous}. 
%
%
%
\begin{gather}
	I_{\text{net}} = I_{\text{s}} + I_{\text{d}} +I_{\text{c}} ={\int_{-1}^{1} {n^{\ast}\myvec{V}} dy}  - {\int_{-1}^{1} {\frac{1}{Pe}\left[\frac{\partial n_{\text{+}}}{\partial x}-\frac{\partial n_{\text{-}}}{\partial x}\right]} dy} - {\int_{-1}^{1} {\frac{1}{Pe}\left[(n_{\text{+}}+n_{\text{-}})\frac{\partial U}{\partial x}\right]} dy} =0
	\label{eq:7}
\end{gather}
where $I_{\text{s}}$, $I_{\text{d}}$ and $I_{\text{c}}$ are the convection (or streaming), diffusion and conduction current densities, respectively. Diffusion current is zero ($I_\text{d}=0$) at the steady state. 
Further, the axial potential gradient ($\partial U/\partial x$) relates with induced electric field strength ($E_\text{x}=-\partial U/\partial x$).


\noindent
[BC-4]:  
The solid walls of the device are no-slip and impermeable (i.e.,  zero velocity in both normal and tangential to the walls, and zero flux density of ionic species normal to the walls), and uniform but asymmetrically charged.
%
%
\begin{gather}
	V_{\text{n}} =0,
	\qquad 
	V\subt =0,
	\qquad
	\myvec{f}_{\text{j}}\cdot \myvec{n}\subb=0,
	\qquad
	(\nabla U\cdot\myvec{n}\subb) = S_\text{i}
	\label{eq:9}
\end{gather}
%
%
%
%
%
where, $V_{\text{n}}$ and $V\subt$ are the normal and tangential components of the wall velocity, respectively, $\myvec{f}_{\text{j}}$ is the flux density of ionic species described by the Einstein relation \citep[refer Eq. A.5 in][]{dhakar2022electroviscous}.
%
%
%

\noindent 
Further, $S_{\text{i}}$ is  the dimensionless surface charge density at the walls (\rev{i = b, t; i.e., $\sut$ and $\sub$ at the top and bottom walls}) of the microfluidic device. The asymmetry of surface charge on the walls is expressed by the surface charge density ratio ($\sur$) as follows.
\begin{gather}
		S_{\text{i}}=\frac{\sigma_{\text{i}}W}{\varepsilon_{\text{0}}\varepsilon\subr U_{\text{c}}}
	\qquad\text{and}\qquad
	\sur=\frac{\sub}{\sut}
	\label{eq:12}
\end{gather}
\noindent
Note that $\sut > 0$ whereas $S_\text{b}\ge 0$. In case of $\sur=0$, only top wall is charged ($\sut > 0$) and the bottom wall is electrically neutral (i.e., zero charge, $S_\text{b}=0$); and both walls are equally charged ($\sut = S_\text{b}$) for $\sur=1$. The top wall charge dominates ($\sut > S_\text{b}$) for $\sur<1$ and bottom wall charge dominates ($S_\text{b} > \sut$) for $\sur>1$.
%
%
%

\noindent
%
The above dimensionless multiphysics coupled mathematical model (\eqns\ref{eq:2} to \ref{eq:9}) has been solved numerically by using the finite element method to obtain the electrokinetic flow fields such as total potential ($U$), ionic species concentration ($n_{\text{j}}$), excess charge ($n^\ast$), induced electric field strength ($E_\text{x}$), velocity ($\myvec{V}$), and pressure ($P$) in the microfluidic device.
%
\section{Numerical approach}
\label{sec:sanp}
%
\noindent 
The EVF model governing equations with relevant boundary conditions (\eqns\ref{eq:2} to \ref{eq:9}) are solved numerically using finite element method (FEM) based solver COMSOL multiphysics software. The COMSOL modules \textit{electrostatic} (es) for total electrical potential, \textit{transport of dilute species} (tds) for ion concentration, and \textit{laminar flow} (spf) for velocity and pressure fields are used to represent the two-dimensional (2-D) fully coupled multiphysics problem. The uniform (except boundary and corners), rectangular mesh  and linear shape function have used to discretize the computational domain. The boundary layer and corner refinement have been implemented in all meshes to account the sharp changes at corners and boundary effects. Integrals in \eqn(\ref{eq:7}) have been solved by using the $intop$ function present in the component definition section of model coupling. A finite element method is used to transform the partial derivatives and partial differential equations (PDEs) to the simultaneous algebraic equations (SAEs). The $1$st order polynomials ($P_\text{p}+P_\text{q}$ with $p=q=1$), i.e., the shape functions with $p^{th}$ and $q^{th}$ order elements, are used for velocity and pressure fields spatial discretization. The methodology involves fully coupled PARDISO linear and Newton's non-linear solvers in which steady-state solution is obtained for the Poisson's, Nernst-Planck and Navier-Stokes equations for fully developed flow in asymmetrically charged contraction-expansion microfluidic device. The steady state solution yields the total electrical potential ($U$), ion concentration ($n_\text{+}$ and $n_\text{-}$), induced electric field strength ($E_\text{x}$), velocity ($\myvec{V}$) and pressure ($P$) fields. 

\noindent
The mesh and domain independence studies are performed to obtain the final accurate results free from the numerical and end effects. The mesh independence test is carried out using three (M$_1$, M$_2$, and M$_3$) mesh in the electroviscous flow through the contraction-expansion microfluidic device. \tab\ref{tab:1m} reports the mesh details and influence of mesh on the pressure drop ($10^{-5}|\Delta P|$) for the range of $K$ and $S_\text{t}$. The pressure drop values are seen to change maximally by 0.1\% for the considered meshes (M$_1$, M$_2$, and M$_3$). An increasing mesh size (i.e., $N_\text{e}$) significantly increases the computational efforts. Hence, the mesh M$_2$ is used in this work to obtain the final results.
\begin{table}[!tb]
	\centering
	\caption{Mesh characteristics and influence of mesh on the pressure drop ($10^{-5}|\Delta P|$) in symmetrically charged ($S_\text{r}=1$)  contraction-expansion microchannel.}\label{tab:1m}
	\scalebox{0.9}
	{
		\begin{tabular}{|c|c|c|c|c|c|c|c|}
			\hline
			\multicolumn{5}{|r|}{Mesh} & M$_1$   & M$_2$    & M$_3$	  \\\hline
			\multicolumn{5}{|r|}{Mesh points per unit length} & 50   & 100   & 150	  \\\hline
			\multicolumn{5}{|r|}{Number of elements ($N_\text{e}$)} & 85550	&	333600	& 744150	  \\\hline
			\multicolumn{5}{|r|}{Degree of freedom (DoF)} & 778139	&	3018814	& 6721989	  \\\hline
			\multicolumn{5}{|r|}{Relative tolerance} & $10^{-5}$	&	$10^{-5}$	&	$10^{-5}$	  \\\hline
			\multicolumn{5}{|c|}{$10^{-5}|\Delta P|$ (at $S_\text{r}=1$) } & \multicolumn{3}{c|}{Error (\%)} 	\\\hline
			$S_\text{t}$ &	$K$ &	M$_1$	&	M$_2$	&	M$_3$	& M$_1$–M$_2$ &	M$_2$–M$_3$ & M$_1$–M$_3$  \\\hline
			4 &	2   & 	1.1697   & 	1.1690	 &   1.1688    &   0.060    & 	 0.018 &  0.078	 \\
			&	4	&	1.1205 	 &	1.1199	 &	 1.1197    &   0.054	&	 0.018 &  0.072	 \\
			&	6   &	1.0967	&	1.0960	 &	 1.0958	   &   0.065	&	 0.018 &  0.083	 \\
			&	8   &	1.0825	&	1.0818	 &	 1.0817	   &   0.066	&	 0.009 &  0.075  \\
			&	20  &	1.0642	&	1.0636	 &	 1.0634	   &   0.057	&	 0.019 &  0.076   \\\hline
			16 &	2   & 	1.4261   & 	1.4250	 &   1.4248    &   0.076    & 	 0.014 &  0.090	 \\
			&	4	&	1.3667 	 &	1.3658	 &	 1.3656    &   0.067	&	 0.015 &  0.081	 \\
			&	6   &	1.2986	&	1.2979	 &	 1.2978	   &   0.055	&	 0.008 &  0.062 \\
			&	8   &	1.2367	&	1.2361	 &	 1.2360	   &   0.049	&	 0.008 &  0.057 \\
			&	20  &	1.0827	&	1.0822	 &	 1.0821	   &   0.047	&	 0.009 &  0.056   \\\hline
		\end{tabular}
	}
\end{table}
Based on the mesh independence study and existing knowledge \citep{davidson2007electroviscous,dhakar2022electroviscous}, the following geometrical and mesh parameters are used for the numerical simulations: 
\\\noindent (i) geometrical parameters: $W= 0.1~\mu$m, $d_\text{c}=0.25$, and $L_\text{u}=L_\text{c}=L_\text{d}=5W$. 
\\\noindent (ii) mesh characteristics: uniform, rectangular mesh M$_2$, 100 grid points per unit length, the total number of grid elements, $N_\text{e}=333600$, degree of freedom, DoF $=3018814$ (refer \rev{\fig\ref{fig:1b} and} \tab\ref{tab:1m}). 
Subsequently, the new results free from ends, and mesh effects are presented and discussed in the next section.
%
%
%
%
\section{Results and discussion}
\noindent 
In this study, the detailed numerical results have been obtained to quantify the electroviscous effects (EVEs) in  pressure-driven flow of symmetric ($1{:}1$) electrolyte liquid through a asymmetrically charged non-uniform (contraction-expansion) slit microfluidic device for the following broader ranges of conditions: inverse Debye length  (\{$2K~|~K\in[1..10]$\});
top wall surface charge density  (\{$2\sut~|~\sut\in[2^1..2^3]$\});
surface charge density ratio (\{$0.25\sur~|~\sur\in[0..8]$\});
Reynolds number ($Re=10^{-2}$); Schmidt number ($\mathit{Sc}=10^3$) and liquid parameter ($\beta=2.34\times10^{-4}$). 

\noindent
The above ranges of conditions are justified \citep{davidson2007electroviscous,bharti2008steady,bharti2009electroviscous,dhakar2022electroviscous} as follows. The low Reynolds number ($Re=10^{-2}$) is considered as microfluidic flows are generally laminar, and the Schmidt number is taken as $\mathit{Sc}=1000$ based on the water properties. The extreme values of inverse Debye length ($K$) account for the strongest and weakest EV effects due to thick (or overlapped) EDL at $K=2$ and very thin EDL at $K=20$.
The surface charge density ($S_\text{i}$) variation from 4 to 16 depicts the practical ranges of the zeta potential ($\zeta$, mV) as $50-100$ (at $K=2$, overlapping EDL) and $12-50$ (at $K=8$, thin EDL) for the uniform geometries. The bulk ion concentration ($n_{0}$) is calculated as per the literature \citep{davidson2007electroviscous}.
Further, the charge density ($S_\text{i}$) corresponds with the practical values of the dimensional charge density ($\sigma_\text{i}$, mC/m$^2$) as $\sigma_\text{i} = m S_\text{i}$ where $m=(2/11)$ in this work. Henceforth, all the parameters and conditions are dimensionless until otherwise mentioned.

\begin{figure}[tb]
	\centering\includegraphics[width=0.85\linewidth]{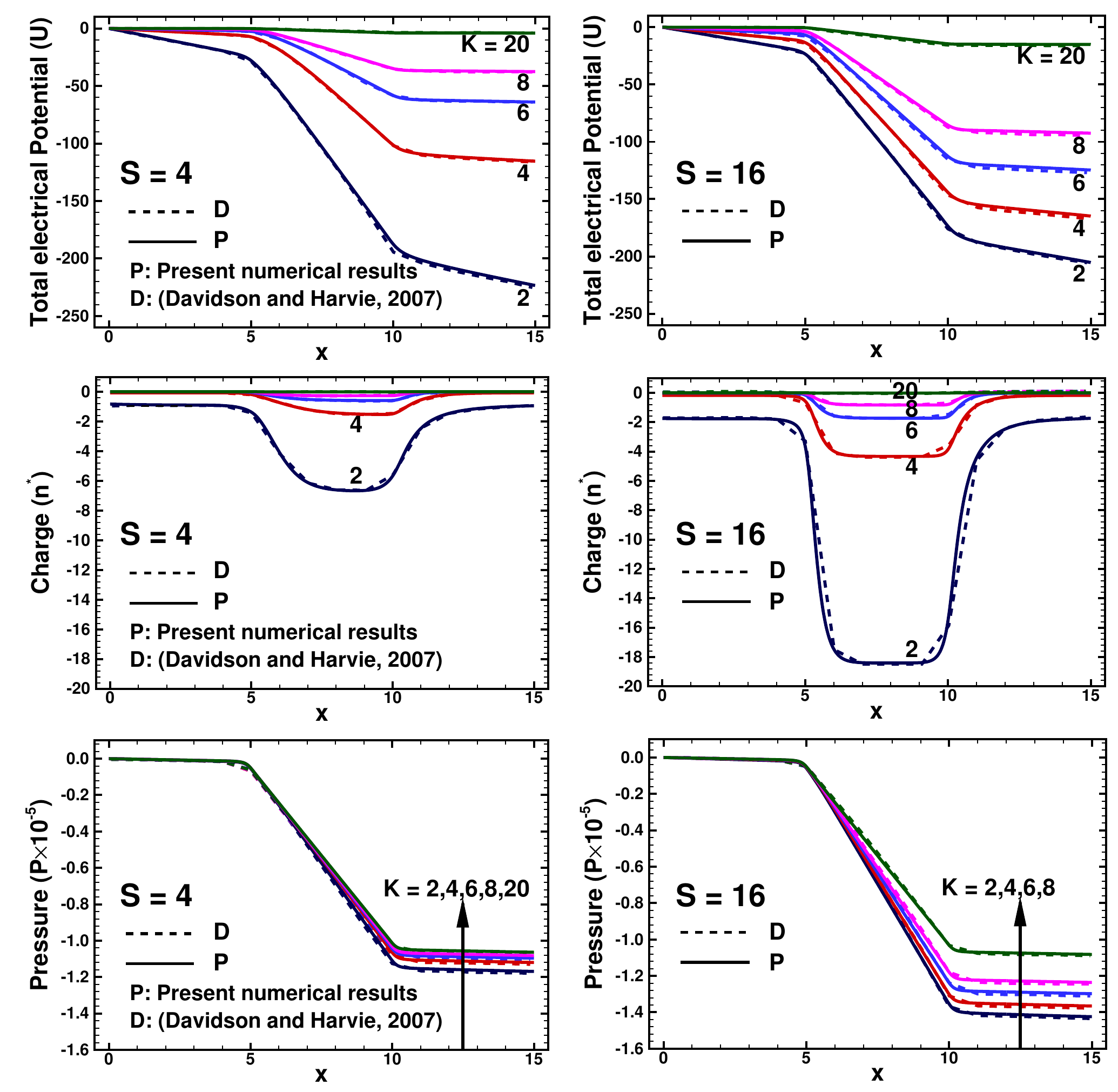}
	\caption{ Comparison of centreline profiles of total potential ($U$), excess charge ($n^\ast$), and pressure ($P$) in the symmetrically charged ($S_\text{r}=1$) contraction-expansion microfluidic device, present results (solid line) with the literature (dashed line) \citep{davidson2007electroviscous}.}\label{fig:comparison}
\end{figure}

\noindent
The modelling approaches used in the present work have been thoroughly validated for the limiting case of electrolyte flow through symmetrically charged ($\sur=1$) non-uniform (contraction-expansion) microfluidic device for the broad ranges of the conditions (see \figs\ref{fig:comparison}, and  \citep[\fig 10, in][]{dhakar2022electroviscous}). 
The comparison of present and literature \citep{davidson2007electroviscous} values has shown an excellent ($\pm$1--2\%) agreement for flow fields (i.e., total potential ($U$), excess charge ($n^\ast$), pressure ($P$), and electroviscous correction factor ($Y$)) in the microfluidic device. Furthermore, none of the results are available for electrolyte flow through asymmetrically charged ($S_\text{r}\neq 1$) surfaces of the microfluidic geometry in the existing literature, and thus direct comparison of the result is not possible. Based on the above comparison and previous experience \citep{davidson2007electroviscous,bharti2008steady,bharti2009electroviscous,Venkat2021,dhakar2022electroviscous}, the present results are accurate and reliable within an excellent ($\pm 1-2\%$) level of accuracy.
\\\noindent 
Subsequently, the new results obtained in this study are presented and discussed in terms of total electrical potential ($U$), excess charge ($n^\ast$), induced electric field strength ($E_\text{x}$), pressure ($P$) and electroviscous correction factor ($Y$) as a function of flow governing parameters ($K$, $\sut$ and $\sur$).
%
\subsection{Total electrical potential}
\label{sec:potential}
\begin{figure}[tb]
	\centering\includegraphics[width=0.9\linewidth]{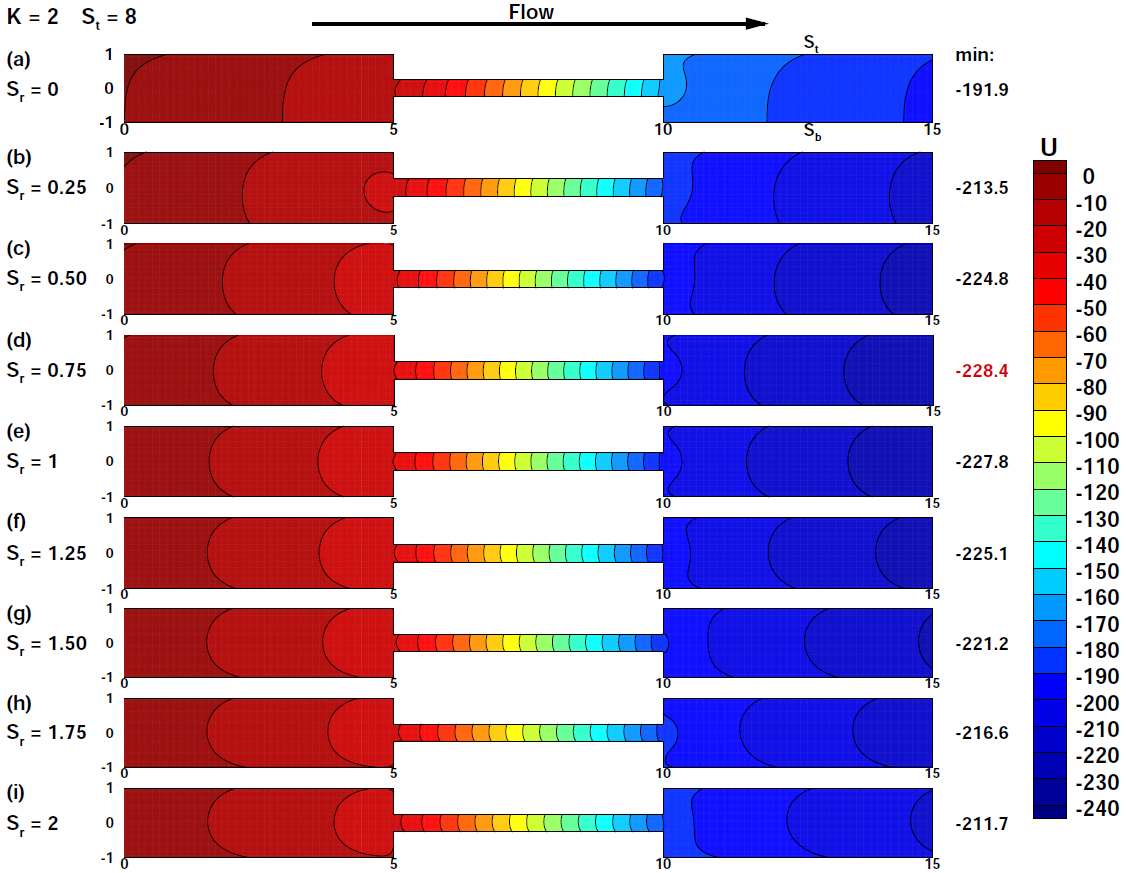}
	\caption{Total electrical potential ($U$) distribution for $K=2$, $\sut=8$ and $0\le \sur\le 2$.}
	\label{fig:2}
\end{figure}

\noindent
The mathematical model suggests that the local distribution of the total electrical potential ($U$) field in the asymmetrically charged slit microfluidic device depends on the flow governing parameters ($K$, $\sut$ and $\sur$). 
Since the potential contour profiles  are qualitatively similar for the ranges of conditions ($2\le K\le 20$, $4\le \sut\le 16$ and $0\le \sur\le 2$), \fig\ref{fig:2} illustrates the representative contours of $U$ for $K=2$, $\sut=8$ and $0\le \sur\le 2$. 
As expected, the electrical potential ($U$) decreases along the length of the device. The contour profiles are symmetric about the horizontal centerline ($x,0$) for the symmetrically charged ($\sur=1$) microfluidic system,  whereas the symmetricity is absent for  $\sur\neq 1$.  For instance, the contour line coincides with the inward normal of  uncharged surface ($S_\text{i}=0$), as expressed in \fig\ref{fig:2}a where the surface charge density of the bottom wall is zero ($\sur=0$, i.e.,  $S_\text{b}=0$). The lateral curving in the contours is appeared from the charged surface toward the core due to the non-zero potential gradient ($\nabla U$) normal to the charged surface, which is equal to the surface charge density (\eqn\ref{eq:9}).  Broadly, the lateral curving varies with increasing $\sur$ from $0$ to $2$ such as lateral curving is symmetric at $\sur=1$, more effective at lower surface charge ratio ($\sur<1$) and less effective at higher surface charge ratio ($\sur>1$). 
The potential gradient is maximum in the contraction compared to the upstream and downstream sections. It is due to the enhancement in the streaming current (and hence reduction in the streaming potential) attributed to increasing convective flow velocity with the reduction in flow area in the contraction. 

\noindent 
Further, the potential ($U$) decreases with an increasing asymmetry of surface charge ($\sur$). It is due to the strengthening of the electrostatic forces near the walls, which increases the excess charge available for transport in EDL. At higher $\sur$, the strengthening of charge attraction retards the convective flow of ions in EDL, therefore, $U$ increases with increasing $\sur$. The minimum value of $U$ is obtained as $-228.4$ at $\sur=0.75$ for $K=2$, $\sut=8$ (as shown in \fig\ref{fig:2}d). 
Furthermore, the total potential ($U$) profiles for limiting condition ($\sur=1$) have shown consistent behavior with literature \citep{davidson2007electroviscous,davidson2008electroviscous,bharti2008steady,bharti2009electroviscous,dhakar2022electroviscous}. 

\begin{figure}[!t]
	\centering\includegraphics[width=1\linewidth]{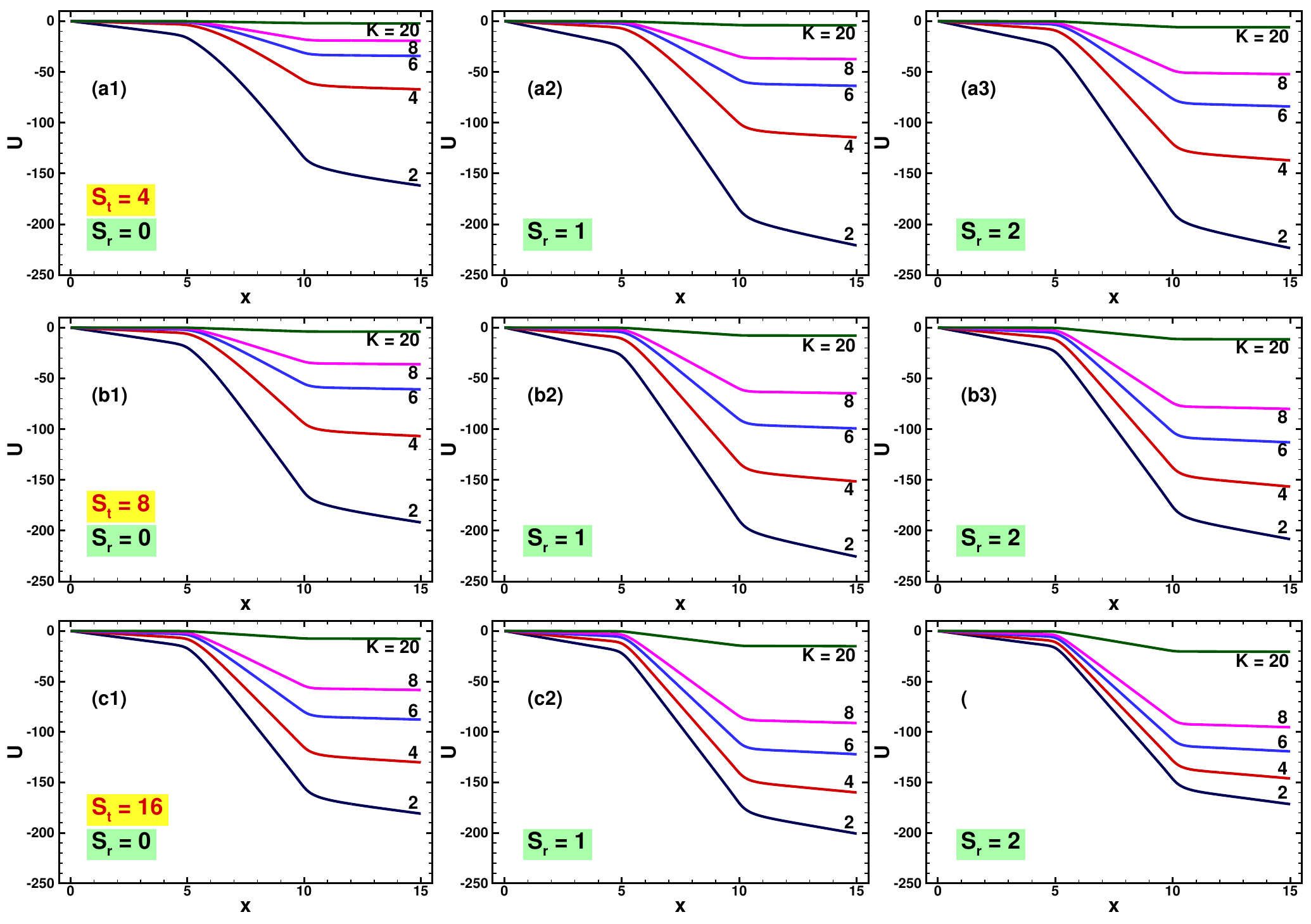}
	\caption{Total electrical potential ($U$) on the centreline ($x,0$) of the asymmetrically charged microfluidic device as a function of dimensionless governing parameters ($K$, $\sut$ and $\sur$).}
	\label{fig:3}
\end{figure}

\noindent
\fig\ref{fig:3} depicts the total electrical potential ($U$) variation on the centreline ($x,0$) of the asymmetrically charged microfluidic device for the considered ranges of conditions. The potential ($U$) decreases along the length ($0\le x\le L$) of a microfluidic device, irrespective of the flow conditions ($K$, $S_\text{i}$).  
The advection of the negative ions (excess charge) available for convection transport enhances due to the positively charged surfaces; thereby, the streaming current ($I_\text{s}$) increases and the streaming potential decreases with increasing $\sur$. The potential drop ($\Delta U$) is highest in the contraction (\fig\ref{fig:3}) as compared to other sections of the device.

\noindent 
The potential ($U$) decreases with decreasing $K$ (\fig\ref{fig:3}) as the thickening of EDL augments the excess charge in the diffuse layer of the EDL for transport in the device.  
The influences of charged surfaces ($\sut$ and $\sur$) on the potential ($U$) are seen to be complex. For instance,  the potential ($U$) decreases with increasing $\sut$ and $\sur$ (as shown in \fig\ref{fig:3}a) but at higher $\sut$ and $\sur$, it increases with increasing $\sut$ and $\sur$ (as shown in \fig\ref{fig:3}c). The maximum change in the potential drop is noted as 197.45\% at the weak electroviscous flow condition ($\sut=4$ and $K=20$), over the ranges of conditions. The qualitative trends for total potential profiles are the same as shown in the literature \citep{davidson2007electroviscous,dhakar2022electroviscous}.
\begin{table}[!b]
	\centering
	\caption{The total electrical potential drop ($|\Delta U|$) on the centreline ($x,0$) over the length ($L$) of the asymmetrically charged microfluidic device. The maximum potential drop (i.e., $|\Delta U|\smax$) for each $K$ and $\sut$ is highlighted with underline.}\label{tab:1}
	\scalebox{0.77}
	{
	\begin{tabular}{|r|r|r|r|r|r|r|r|r|r|r|}
		\hline
		$\sut$	&	$K$	&	\multicolumn{9}{c|}{$|\Delta U|$}	\\\cline{3-11} 
		&		&	$\sur=0$	&	$\sur=0.25$	& $\sur=0.50$ &	$\sur=0.75$	& $\sur=1$ & $\sur=1.25$ & $\sur=1.50$ & $\sur=1.75$ & $\sur=2$ \\\hline 
		0   & 	$\infty$    & 	0	    & 0	        & 0          & 	0	    & 0         & 	0	& 0	&0	        &0 \\\hline 
		4	&	2	&	161.9300	&	186.8400	&	203.8600	&	214.6300	&	220.8400	&	223.9300	&	\htxt{224.9700}	&	224.6700	&	223.4600	\\
&	4	&	67.1180	&	81.5850	&	94.3250	&	105.2500	&	114.4100	&	121.9800	&	128.1600	&	133.1400	&	\htxt{137.1200}	\\
&	6	&	34.3130	&	42.3930	&	50.0260	&	57.1380	&	63.6750	&	69.6200	&	74.9710	&	79.7450	&	\htxt{83.9730}	\\
&	8	&	19.3580	&	24.0910	&	28.6990	&	33.1450	&	37.4020	&	41.4440	&	45.2590	&	48.8320	&	\htxt{52.1640}	\\
&	20	&	1.9993	&	2.4996	&	2.9990	&	3.4969	&	3.9929	&	4.4864	&	4.9767	&	5.4638	&	\htxt{5.9470}	\\\hline
8	&	2	&	191.8400	&	212.8500	&	223.4600	&	\htxt{226.5600}	&	225.6000	&	222.5200	&	218.3100	&	213.4900	&	208.3500	\\
&	4	&	106.8800	&	124.4000	&	137.1200	&	145.8400	&	151.4900	&	154.8500	&	156.5500	&	\htxt{157.0100}	&	156.5900	\\
&	6	&	60.7930	&	73.3840	&	83.9730	&	92.5630	&	99.3190	&	104.5000	&	108.3500	&	111.1300	&	\htxt{113.0300}	\\
&	8	&	36.1470	&	44.5300	&	52.1640	&	58.9300	&	64.7870	&	69.7570	&	73.9020	&	77.3000	&	\htxt{80.0430}	\\
&	20	&	3.9648	&	4.9594	&	5.9469	&	6.9228	&	7.8842	&	8.8269	&	9.7472	&	10.6430	&	\htxt{11.5120}	\\\hline
16	&	2	&	180.9200	&	202.5900	&	\htxt{208.3500}	&	206.1700	&	200.7000	&	193.8000	&	186.3600	&	178.8500	&	171.5300	\\
&	4	&	130.1300	&	147.4600	&	156.5900	&	\htxt{160.0300}	&	159.9600	&	157.8100	&	154.4400	&	150.4100	&	146.0600	\\
&	6	&	87.7650	&	102.9000	&	113.0300	&	119.0500	&	122.0800	&	\htxt{123.0300}	&	122.5900	&	121.2300	&	119.2800	\\
&	8	&	58.4520	&	70.6530	&	80.0430	&	86.7330	&	91.1610	&	93.8360	&	95.2000	&	\htxt{95.6150}	&	95.3500	\\
&	20	&	7.6769	&	9.6215	&	11.5120	&	13.3210	&	15.0260	&	16.6100	&	18.0650	&	19.3870	&	\htxt{20.5780}	\\\hline		
	\end{tabular}
}
\end{table}
\\
\noindent 
Subsequently, \tab\ref{tab:1} quantitatively summarizes the total electrical potential drop (\mm{|\Delta U|}) on the centreline (\mm{x,0}) over the length (\mm{L}) of the asymmetrically charged microfluidic device for the broader ranges of parameters (\mm{2\le K\le 20}, \mm{4\le \sut\le 16} and \mm{0\le \sur\le 2}). The maximum potential drop (\mm{|\Delta U|\smax}) for \mm{0\le \sur\le 2}  at each combination of \mm{K} and \mm{\sut} is also highlighted as the underlined data. Broadly, the potential drop (\mm{|\Delta U|})  decreases with increasing \mm{K} (i.e., EDL thinning), for the given surface charge density (\mm{S_\text{i}}); this relative impact of \mm{K} is lessening with increasing \mm{S_\text{i}}. For instance, \mm{|\Delta U|} reduces by 98.19\%, 96.51\% and 92.51\% for \mm{S_\text{t}=4}, 8 and 16, respectively, when \mm{K} increases from 2 to 20 at a symmetrically charged (\mm{\sur=1}) condition  \citep{dhakar2022electroviscous}. 
The maximum change in \mm{|\Delta U|} with decreasing charge asymmetry (\mm{\sur<1}), observed for \mm{K=20}, is reducing with decreasing \mm{K}. For instance, \mm{|\Delta U|} reduces by (26.68\%, 14.96\% and 9.86\%) and (49.93\%, 49.71\% and 48.91\%) for \mm{\sut = } 4, 8 and 16, respectively at \mm{K=2} and 20 with decreasing charge asymmetry from 1 to 0 (\mm{\sur< 1}). 

\noindent On the other hand, the corresponding changes in \mm{|\Delta U|}  with increasing charge asymmetry from 1 to 2 (\mm{\sur> 1}) are (1.19\%, -7.65\% and -14.53\%) and (48.94\%, 46.01\% and 36.95\%) for \mm{K=2} and 20, respectively. The overall changes in \mm{|\Delta U|} with the variation in charge asymmetry from 0 to 2 (\mm{0\le \sur\le 2}) are noted as (38\%, 8.61\% and -5.19\%) and (197.45\%, 190.36\% and 168.05\%) for \mm{\sut=4}, 8 and 16 at \mm{K=2} and 20, respectively.
It is further noted that the change in \mm{|\Delta U|} is greater  at \mm{\sut=4} as compared to \mm{\sut=8} and 16, irrespective of the other parameters (\mm{K} and \mm{\sur}) (as shown in \tab\ref{tab:1}).
The drop \mm{|\Delta U|} shows a decrease followed by an increase with increasing \mm{\sut} (as shown in \fig\ref{fig:3} and \tab\ref{tab:1}).
It is because the electrostatic forces strengthen with increasing \mm{\sut} and which impedes the excess ions flow in the EDL in the direction of pressure-driven flow, and therefore, streaming current decreases and hence streaming potential.
%
\begin{figure}[t]
	\centering
	\subfigure[\mm{|\Delta U|\smax}]{\includegraphics[width=0.49\linewidth]{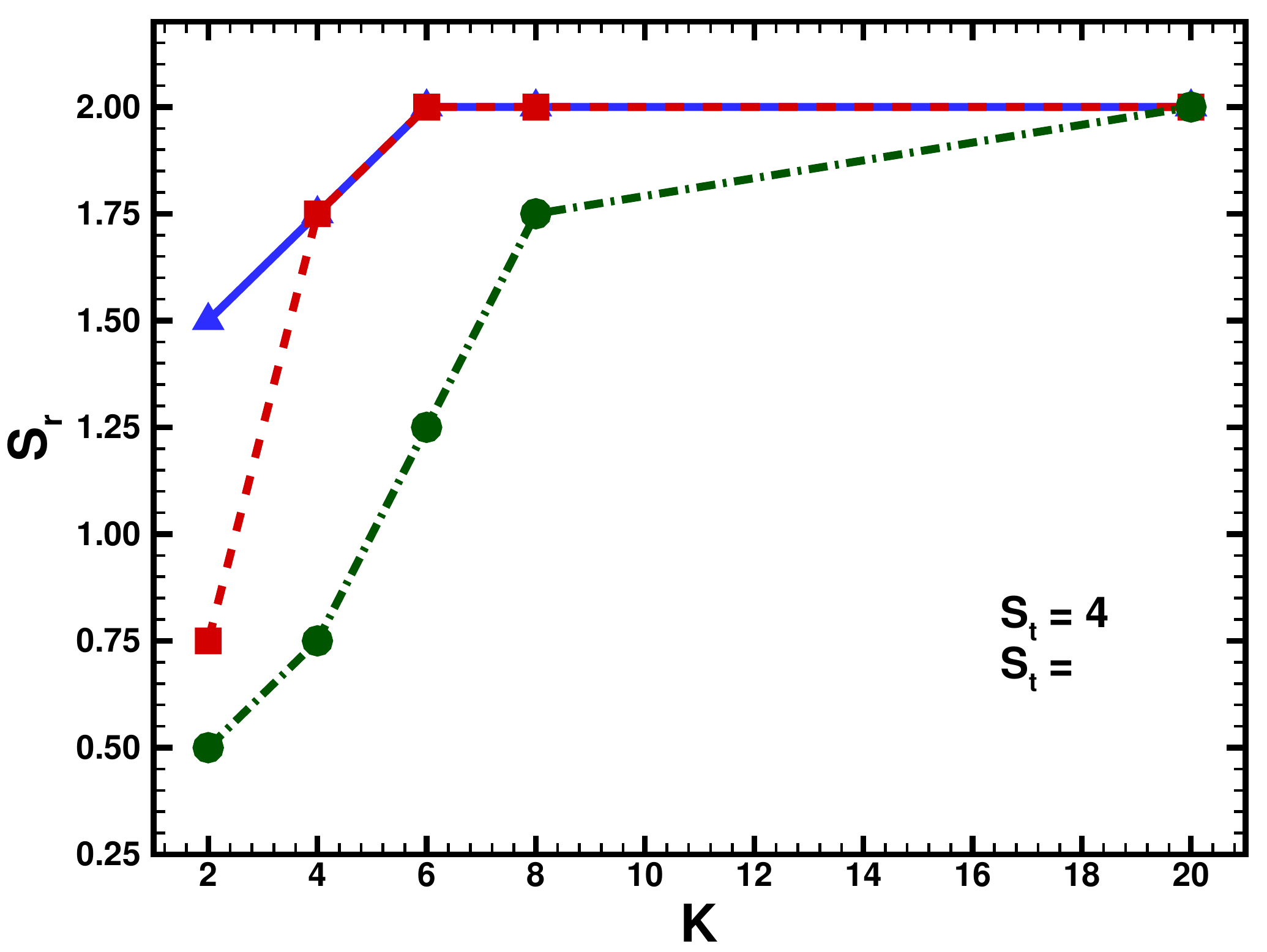}\label{fig:13a}}
	\subfigure[\mm{E_\text{x,max,h}}]{\includegraphics[width=0.49\linewidth]{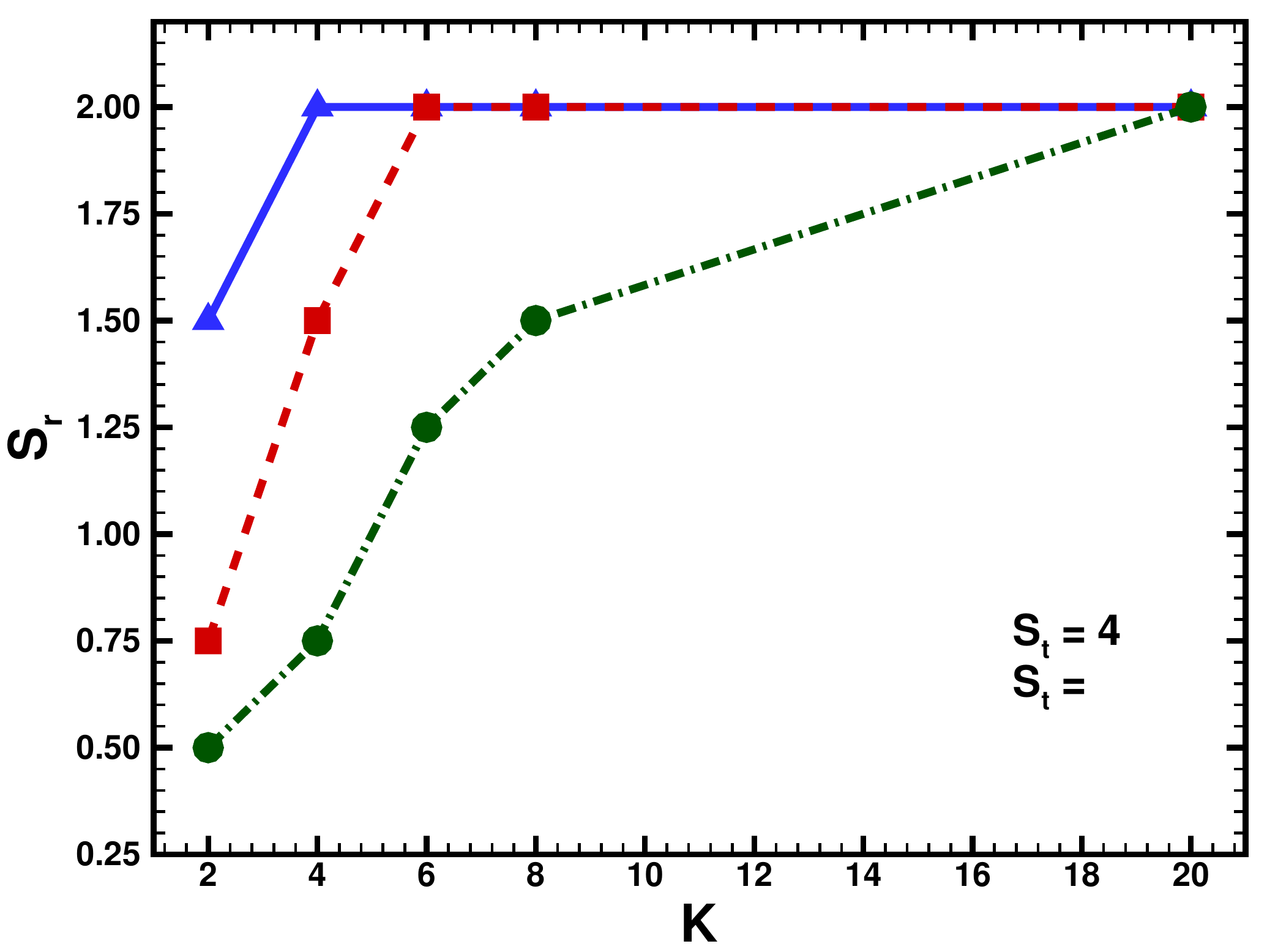}\label{fig:13b}}
	\caption{Conditional (\mm{K, \sut, \sur}) maps representing the highest values of \mm{|\Delta U|}, and \mm{E_\text{x,max}}.}
	\label{fig:13}
\end{figure} 
\\\noindent
\fig\ref{fig:13a} maps the conditions (\mm{K, \sut, \sur}) which yield the maximum potential drop (\mm{|\Delta U|\smax}). The symmetrically charged (\mm{\sur=1}) case yields \mm{\Delta U_\text{min}} at \mm{\sut=8}, as compared to \mm{\sut=4} and 16, due to the overlapping EDL at \mm{\sut=8}. However, except at \mm{K=20} where negligible thick EDL, \mm{|\Delta U|\smax} is obtained at lower \mm{\sur<2} with increasing \mm{\sut} (as shown in \tab\ref{tab:1}).
The shift in \mm{|\Delta U|\smax} with \mm{\sur} is more significant at the lower \mm{K} and smaller at the higher \mm{K} (as shown in \fig\ref{fig:13a}) because \mm{K} is inversely proportional to the EDL thickness. 
For instance, \mm{|\Delta U|\smax} at \mm{K=2} is noted as 224.97 (at \mm{\sur=1.5}), 226.56 (at \mm{\sur=0.75}), and 208.35 (at \mm{\sur=0.5}) for \mm{\sut=4}, 8 and 16, respectively. The values of \mm{|\Delta U|\smax} at \mm{K=8} is noted as 52.164 (at \mm{\sur=2}), 80.043 (at \mm{\sur=2}), and 95.615 (at \mm{\sur=1.75}).
{Thus, at lower \mm{K=2}, EDL overlaps at \mm{\sur=1.50} and \mm{\sut=4}; on the other hand, even at higher \mm{K=8}, EDL overlaps at higher \mm{\sur=1.75} and \mm{\sut=16}, as shown in \fig\ref{fig:13a}. It is due to stronger electrostatic forces close to the device walls impedes the convective flow of ions in the microchannel at higher \mm{\sur} and \mm{\sut}, decreases streaming current and hence streaming potential (as shown in \fig\ref{fig:3} and \tab\ref{tab:1}).}
\begin{figure}[t]
	\centering\includegraphics[width=1\linewidth]{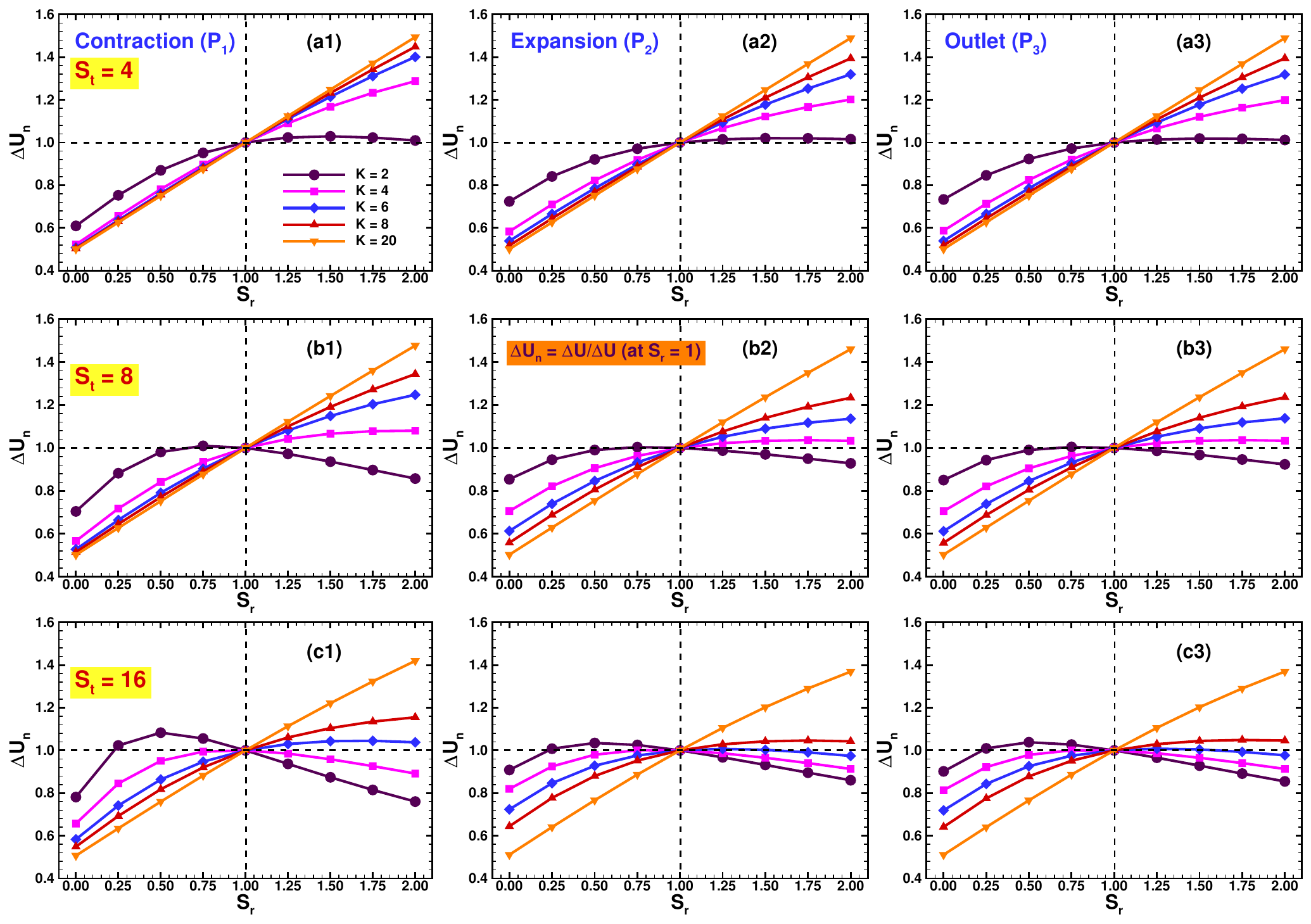}
	\caption{Normalized potential drop (\mm{\Delta U_\text{n}}) variation on contraction ($P_1$, 1st column), expansion ($P_2$, 2nd column), and outlet ($P_3$, 3rd column)  points on the centreline of the microfluidic device with $\sur$ at \mm{K=2} to 20 and $\sut=4$ to $16$.}
	\label{fig:3a}
\end{figure} 
\\
\noindent
Further, the relative influence of charge asymmetry  (\mm{\sur}) is understood by normalizing the potential drop (\mm{\Delta U}) for asymmetrically charged (\mm{\sur\neq 1}) by that at the reference condition of symmetrically charged (`ref' or \mm{\sur = 1}) microfluidic device for other identical parameters (\mm{K, \sut}) as follows,
\begin{gather}
	\Phi_\text{n} = \frac{\Phi}{\Phi_{\text{ref}}} = 
	\left.\frac{\Phi(\sur)}{\Phi(\sur=1)}\right|_{K, \sut}
	\qquad\text{where}\qquad \Phi = (\Delta U, n^{\ast}, E_{\text{x}}, \Delta P)
	\label{eq:Phin}
\end{gather}
\fig\ref{fig:3a} depicts the normalized potential drop (\mm{\Delta U_\text{n}}) variation on the contraction (P$_1$), expansion (P$_2$), and outlet (P$_3$) points on the centreline (refer \fig\ref{fig:1}) of the microfluidic device with flow governing parameters (\mm{K, \sut, \sur}). 
The normalized value (\mm{\Delta U_\text{n}}) increases with increasing $\sut$ and $\sur$ at all the points (P$_1$, P$_2$, P$_3$) (\fig\ref{fig:3a}). The curves cross over at \mm{S\subr=1}, i.e., \mm{\Delta U_\text{n}} decrease for decreasing charge asymmetry (\mm{\sur<1}), and vice-versa for increasing charge asymmetry (\mm{\sur>1}) with increasing \mm{K} (or EDL thinning).  
For instance, \mm{\Delta U_\text{n}} at \mm{K=2} reduces by (39.05\%, 27.61\%, 26.68\%) and (21.89\%, 9.19\%, 9.86\%) at points (P$_1$, P$_2$, P$_3$) for \mm{\sut=4} and 16, respectively, with decreasing charge asymmetry (\mm{\sur<1}); the corresponding reduction in \mm{\Delta U_\text{n}} at \mm{K=20} is noted as (49.96\%, 49.93\%, 49.93\%) and (49.30\%, 48.89\%, 48.91\%) for \mm{\sut=4} and 16, respectively.

\noindent
On the other hand, \mm{\Delta U_\text{n}} changes at  \mm{K=2}  by (0.97\%, 1.53\% and 1.19\%) and (-23.99\%, -14.01\%, -14.53\%) at points (P$_1$, P$_2$, P$_3$) for \mm{\sut=4} and 16, respectively, with increasing charge asymmetry from 1 to 2 (\mm{\sur>1}); the corresponding {increment} in \mm{\Delta U_\text{n}} at \mm{K=20} is noted as (49.38\%, 48.93\%, 48.94\%) and (41.94\%, 36.89\%,  36.95\%) for \mm{\sut=4} and 16, respectively.
Thus, \mm{\Delta U_\text{n}} weakens with decreasing \mm{K} due to the thickening of EDL, which also overlaps at small \mm{K}.
Further, the maximum change in \mm{\Delta U_\text{n}} is obtained (\fig\ref{fig:3a}) at contraction point (P$_1$) as compared to the others (P$_2$, P$_3$)  with flow governing parameters (\mm{K,\sut,\sur}) because clustering of excess charge (EDL overlapped) obtains due to reduction in the cross-section area in the contraction section of the microfluidic device.
%
\\
\noindent The predictive correlation for the potential drop (\mm{\Delta U}, \tab\ref{tab:1}) over the centreline (\mm{x,0}) of the contraction-expansion asymmetrically charged microfluidic slit as a function of the dimensionless parameters (\mm{K,\sut,\sur, d_\text{c}=0.25}) is expressed as follows.
\begin{gather}
	\Delta U = \alpha_1\Delta U_\text{ref} \label{eq:delUP}
	\\
	 \alpha_1 = \beta_1 + (\beta_2  + \beta_4 \sur) \sur+ (\beta_3  + \beta_5 \gamma) \gamma+ 
	 (\beta_6 + \beta_7 \sur \gamma^{-3}) \sur \gamma;
\nonumber
\\ 
	\beta_{\text{i}} = \sum_{{j}=1}^5 M_{\text{ij}} X^{({j}-1)};
	\quad X = K^{-1}; \quad  \gamma = \sut^{-1};
	\quad 1\le i\le 7 \nonumber
\end{gather}
where \mm{\Delta U_\text{ref}} is the potential drop \citep[\eqn (14) in ref.][]{dhakar2022electroviscous} for liquid flow through symmetrically charged (ref: \mm{\sur=1}) contraction-expansion slit.
The correlation coefficients ($M_\text{ij}$, \eqn\ref{M:delU}) are statistically obtained by performing the non-linear regression analysis using the DataFit (version 9.0, free trial) for 135 data points with 
(\mm{\delta\smin, \delta\smax, \delta\avg,R^2}) as (-7.24\%, 3.99\%, -1.65\%, 99.91\%)
 for the ranges of the conditions explored herein. Here, \mm{\delta} and \mm{R^2} are the deviation between  predicted (\eqn\ref{eq:delUP}) and numerical (\tab\ref{tab:1}) values, and the coefficient of determination.
\begin{gather}
M = {\begin{bmatrix}
		0.3106 & 5.3177 & -10.744 & -7.1506 & 25.967 \\
		0.3953 & 5.1061 & -57.749 & 187.43 & -188.89 \\
		2.9865 & -97.822 & 576.57 & -1360.7 & 1131.8 \\
		0.0226 & -0.5801 & -2.6543 & 16.496 & -19.888 \\
		-8.2218 & 282.04 & -1927 & 5044.8 & -4498.9 \\
		0.0527 & -8.3957 & 162.99 & -615.27 & 658.82 \\
		0.0005 & -0.01913 & 0.1483 & -0.432 & 0.4152
	\end{bmatrix}} \label{M:delU}
\end{gather}
%
%
Further, the Poisson's equation (\eqn\ref{eq:2}) relates the total electrical potential (\mm{U}) with excess charge (\mm{n^\ast}). Thus, the subsequent section analyses the excess charge distribution (\mm{n^\ast}) for broader ranges of parameters ($K$, $\sut$, $\sur$).
%
\subsection{Excess charge}
\label{sec:charge}
In this study, the excess charge (\mm{n^\ast}, \eqn\ref{eq:2}) distribution has been analyzed by scaling \mm{n^\ast} as follows: \mm{n^{\ast\ast}=(n^\ast - {n^\ast}\smin)/(n^\ast\smax - n^\ast\smin)}, where subscripts `max' and `min' represent for maximum and minimum values for the given condition. While the excess charge is found to be negative (\mm{n^\ast < 0}) in the positively charged microfluidic device, the scaled charge is positive (\mm{0\le n^{\ast\ast}\le 1}) for all conditions.

\begin{figure}[t]
	\centering\includegraphics[width=1\linewidth]{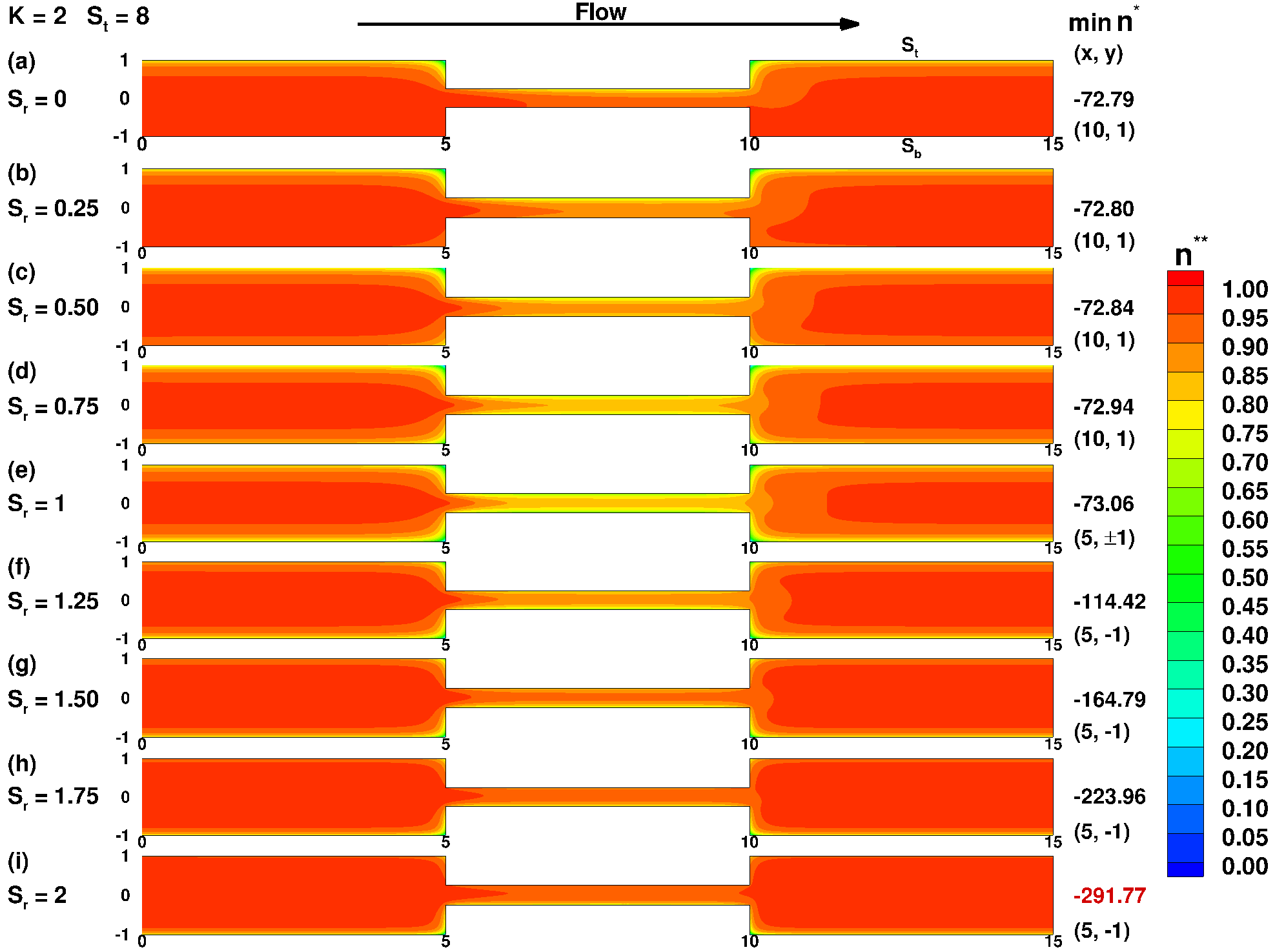}
	\caption{Scaled excess charge (\mm{n^{\ast\ast}}) distribution as a function of \mm{\sur} at \mm{\sut=8} and \mm{K=2}.}
	\label{fig:4}
\end{figure}

\noindent
\fig\ref{fig:4} depicts the scaled excess charge (\mm{n^{\ast\ast}}) distribution, with the minimum \mm{n^\ast} and its location (\mm{x,y}), in the asymmetrically charged microfluidic device for broader ranges of surface charge asymmetry (\mm{0\le \sur\le 2}) at fixed \mm{K=2} and \mm{\sut=8}.  The contours for other ranges of parameters (\mm{2\le K\le 20}, \mm{4\le \sut\le 16}, \mm{0\le\sur\le 2}) are qualitatively same and thus not presented here.
As expected, the mid-plane symmetry in the charge (\mm{n^{\ast\ast}}) contours is lost with increasing/decreasing value of \mm{\sur} from 1. 
In general, the clustering of the excess charge ($n^{\ast\ast}$) is obtained near the positively charged surfaces of the microfluidic device. Further, the dense clustering of charge is depicted in the contraction section, compared to the upstream/downstream section, because of the reduction in the effective flow cross-section area. 

\noindent The charge (\mm{n^{\ast\ast}}) shows complex dependency on the governing parameters (\mm{K,\sut,\sur}). For instance, the scaled charge (\mm{n^{\ast\ast}}) decreases with increasing \mm{\sur} (\fig\ref{fig:4}) because of increasing counter (i.e., negative) ions in the EDL with enhancement in the electrostatic forces near the surfaces.
Intriguingly, \mm{{n^\ast}\smin} is obtained at the top-left corner (10, 1) of the expansion for \mm{\sur < 1}, at both upper/lower-left corners (5, \mm{\pm}1) of the contraction for \mm{\sur =1}, and that at the lower-left corner (5, -1) of the contraction for \mm{\sur >1}  for fixed values of \mm{K=2} and \mm{\sut=8}.  The \mm{{n^\ast}\smin (x,y)}  for \mm{K=2} and \mm{\sut=8} is recorded to reduce from -72.79 (10, 1) to -291.77 (5, -1) with change in \mm{0\le \sur\le 2} (\fig\ref{fig:4}).
\begin{figure}[t]
	\centering\includegraphics[width=1\linewidth]{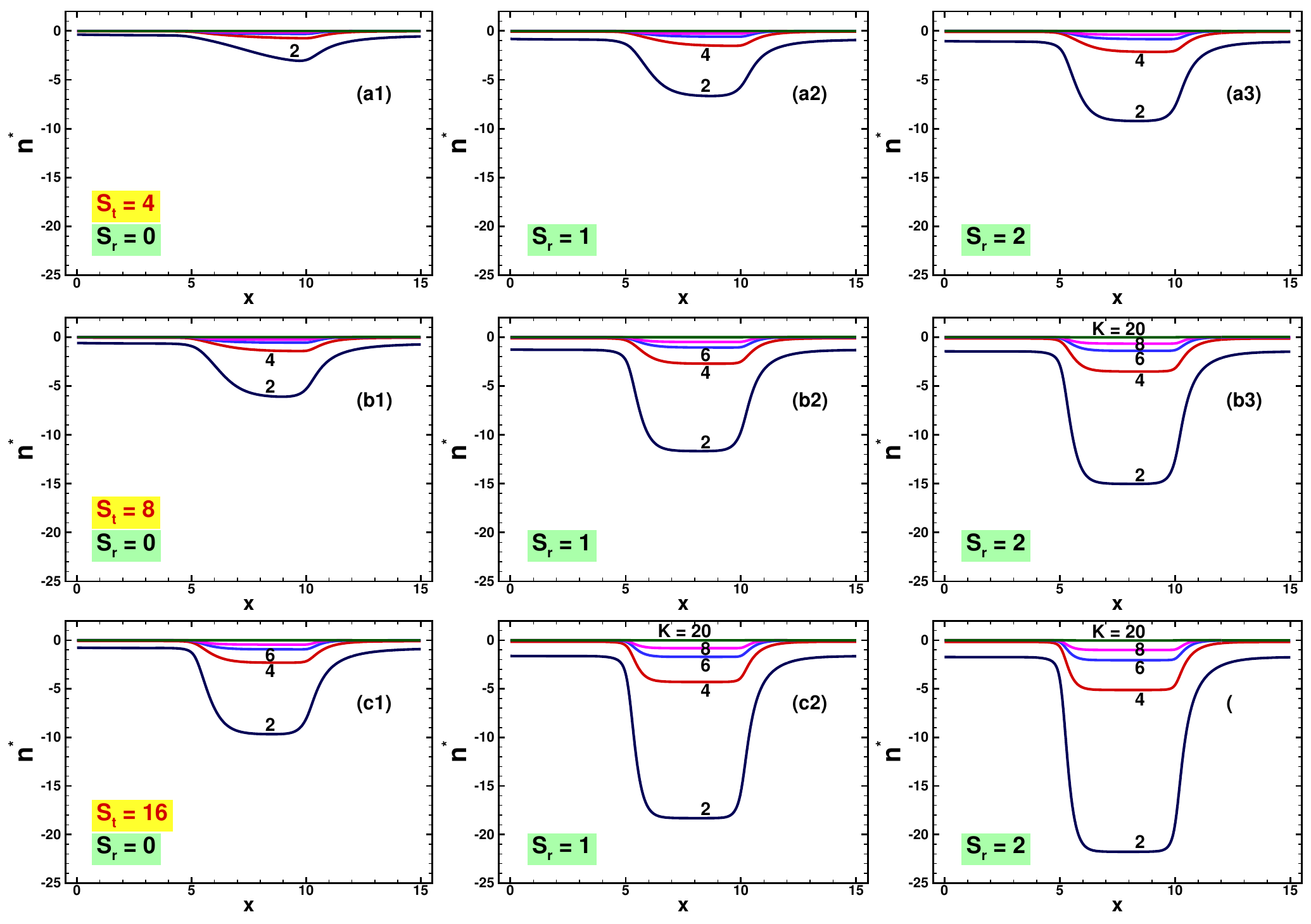}
	\caption{Excess charge (\mm{n^{\ast}}) distribution on the centreline (\mm{x,0}) of the asymmetrically charged microfluidic device as a function of dimensionless governing parameters (\mm{K,\sut,\sur}).}
	\label{fig:5}
\end{figure} 
\\\noindent
Subsequently,  the excess charge (\mm{n^\ast}) variation on the centreline (\mm{x,0}) of the asymmetrically charged microfluidic device with governing parameters (\mm{K,\sut,\sur}) is depicted in \fig\ref{fig:5}. As observed, the excess charge is negative (\mm{n^\ast < 0}) and has shown strong dependence on the flow parameters (\mm{K,\sut,\sur}).
In general, the excess charge (\mm{n^\ast}) is equal and consistent at the center points of the inlet (0, 0) and outlet (L, 0) of the microfluidic device (\fig\ref{fig:5}). It is significantly smaller in the contraction, compared to the upstream/downstream section  because of enhanced clustering of excess charge with reduction in the flow cross-section area, under otherwise identical conditions. 
%
%
\begin{table}
\centering
\caption{Minimum value of excess charge ($n^\ast\smin$) on the centreline of the asymmetrically charged microfluidic device (underlined data shows lowest $n^\ast\smin$ for given $K$ and $\sut$).}\label{tab:2}
\scalebox{0.8}
{
	\begin{tabular}{|r|r|r|r|r|r|r|r|r|r|r|}
		\hline
		$\sut$	&	$K$	&	\multicolumn{9}{c|}{$n^\ast\smin$}	\\\cline{3-11}
		&		&	$\sur=0$	&	$\sur=0.25$	& $\sur=0.50$ &	$\sur=0.75$	& $\sur=1$ & $\sur=1.25$ & $\sur=1.50$ & $\sur=1.75$ & $\sur=2$  \\\hline
		0	&	$\infty$	&	0	& 0		& 0		& 0		& 0 	& 0  &  0  &  0   &  0	\\\hline
		4	& 2	& -3.0562	& -4.1176	& -5.0543	& -5.8922	& -6.6593	& -7.3694	& -8.0297	& -8.6447	& \htxt{-9.2178} \\
		& 4	& -0.7520	& -0.9544	& -1.1540	& -1.3459	& -1.5275	& -1.6977	& -1.8568	& -2.0052	& \htxt{-2.1437} \\
		& 6	& -0.3006	& -0.3757	& -0.4493	& -0.5210	& -0.5901	& -0.6564	& -0.7195	& -0.7795	& \htxt{-0.8362} \\
		& 8	& -0.1353	& -0.1690	& -0.2023	& -0.2350	& -0.2669	& -0.2979	& -0.3279	& -0.3568	& \htxt{-0.3846} \\
		& 20	& -0.0027	& -0.0033	& -0.0040	& -0.0047	& -0.0053	& -0.0060	& -0.0066	& -0.0073	& \htxt{-0.0079} \\\hline
		8	& 2	& -6.1007	& -7.7443	& -9.2178	& -10.5220	& -11.6670	& -12.6720	& -13.5550	& -14.3330	& \htxt{-15.0200} \\
		& 4	& -1.4315	& -1.8036	& -2.1437	& -2.4480	& -2.7178	& -2.9564	& -3.1673	& -3.3542	& \htxt{-3.5202} \\
		& 6	& -0.5650	& -0.7047	& -0.8362	& -0.9570	& -1.0663	& -1.1645	& -1.2524	& -1.3308	& \htxt{-1.4011} \\
		& 8	& -0.2589	& -0.3231	& -0.3846	& -0.4424	& -0.4959	& -0.5451	& -0.5899	& -0.6305	& \htxt{-0.6673} \\
		& 20	& -0.0053	& -0.0066	& -0.0079	& -0.0092	& -0.0105	& -0.0118	& -0.0130	& -0.0143	& \htxt{-0.0154} \\\hline
		16	& 2	& -9.6773	& -12.6400	& -15.0200	& -16.8700	& -18.3160	& -19.4630	& -20.3890	& -21.1480	& \htxt{-21.7790} \\
		& 4	& -2.3184	& -2.9783	& -3.5202	& -3.9509	& -4.2925	& -4.5661	& -4.7882	& -4.9710	& \htxt{-5.1236} \\
		& 6	& -0.9403	& -1.1888	& -1.4011	& -1.5744	& -1.7145	& -1.8280	& -1.9208	& -1.9976	& \htxt{-2.0619} \\
		& 8	& -0.4519	& -0.5660	& -0.6673	& -0.7533	& -0.8247	& -0.8837	& -0.9324	& -0.9731	& \htxt{-1.0073} \\
		& 20	& -0.0103	& -0.0129	& -0.0154	& -0.0179	& -0.0203	& -0.0225	& -0.0246	& -0.0266	& \htxt{-0.0284} \\\hline
	\end{tabular}
}
\end{table}
\\
\noindent \tab\ref{tab:2} subsequently comprises the minimum excess charge ($n^\ast\smin$) on the centreline (\mm{x,0}) of the asymmetrically charged device for the broader ranges of parameters (\mm{K, \sut, \sur}).  The minimum excess charge (\mm{n^\ast\smin}) shows the complex dependence on \mm{K, \sut, \sur}. The trends for excess charge ($n^\ast\smin$) are qualitatively consistent with the literature \citep{davidson2007electroviscous,dhakar2022electroviscous}. It decreases with decreasing \mm{K} and increasing of \mm{\sut} and \mm{\sur} (refer \fig\ref{fig:5} and \tab\ref{tab:2}) as the enhanced electrostatic forces near the device walls increase the excess charge (negative ions) in the EDL.  Further, the lowest \mm{n^\ast\smin} is achieved (\tab\ref{tab:2}) at the largest \mm{\sur=2} irrespective of \mm{K, \sut} values.
The minimum excess charge (\mm{n^\ast\smin}) increases with increasing \mm{K}, or EDL thinning, and approaching zero at higher \mm{K>20}, irrespective of \mm{\sut} and \mm{\sur} (\tab\ref{tab:2}).  
The maximum variation in \mm{n^\ast\smin} with \mm{\sur} is obtained at \mm{K=2}. For instance, \mm{n^\ast\smin} reduces for ($\sut=$ 4, 8, 16)  by (54.11\%, 47.71\%,  47.16\%) and  (49.96\%, 49.85\%, 49.42\%)  at \mm{K=2} and 20 with decreasing \mm{\sur} from 1 to 0.
The corresponding changes in \mm{n^\ast\smin} with increasing \mm{\sur} from 1 to 2 are obtained as (38.42\%, 28.74\% and 18.91\%) and (49.15\%, 46.84\%, 39.76\%) at \mm{K=2} and 20, respectively.
The overall enhancement in \mm{n^\ast\smin} with increasing charge asymmetry (\mm{\sur} from 0 to 2) is obtained as (201.61\%, 146.20\%, 125.05\%) and (198.08\%, 192.79\%, 176.33\%) for (\mm{\sut=} 4, 8, 16) at \mm{K=2} and 20, respectively.  Further, the change in  \mm{n^\ast\smin} is more significant at \mm{\sut=4} as compared to \mm{\sut=8} and 16, under otherwise other identical conditions (\mm{K, \sur}).  It is because more counter ions attracted to the surface, at higher \mm{\sut}, reduce the effective excess charge in the EDL, and therefore, the change is greater in \mm{n^\ast\smin} at \mm{\sut=4} (as shown in \tab\ref{tab:2}). 
\begin{figure}[h]
	\centering\includegraphics[width=1\linewidth]{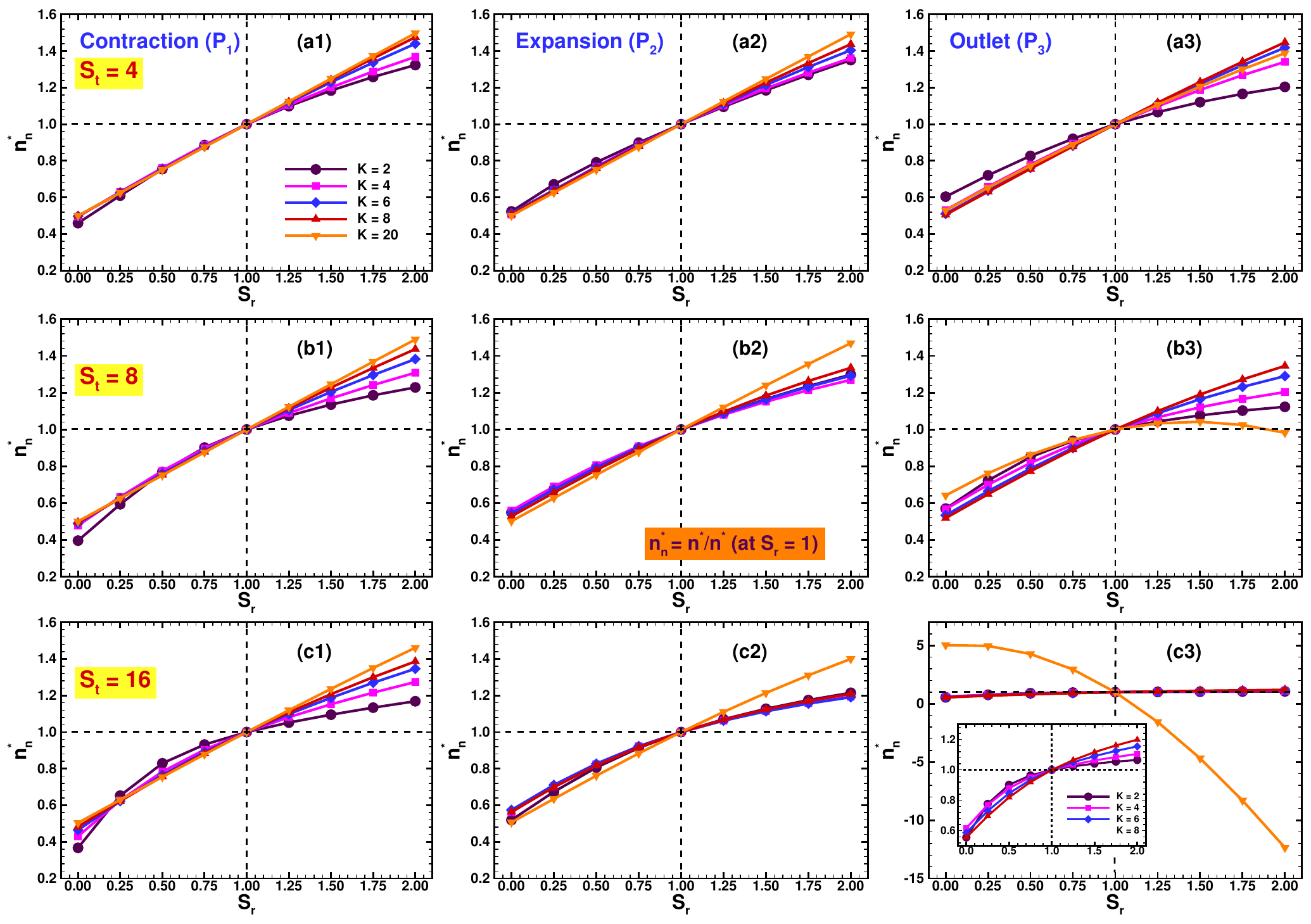}
	\caption{Normalized excess charge ($n^\ast_\text{n}$) variation on contraction ($P_1$, 1st column), expansion ($P_2$, 2nd column), and outlet ($P_3$, 3rd column) centreline locations of the microfluidic device with $\sur$ at $K=2$ to $20$ and $\sut=4$ to $16$.}
	\label{fig:3b}
\end{figure} 
\\
\noindent
Further, \fig\ref{fig:3b} depicts the normalized excess charge (\mm{n^\ast_\text{n}}, \eqn\ref{eq:Phin}) variation on the contraction (P$_1$), expansion (P$_2$), and outlet (P$_3$) points on the centreline (refer \fig\ref{fig:1}) of the microfluidic device with flow governing parameters (\mm{K, \sut, \sur}). 
The normalized excess charge (\mm{n^\ast_\text{n}}) decreases with increasing \mm{K} (or EDL thinning) for decreasing charge asymmetry (\mm{\sur<1}) but increases with increasing \mm{K} for increasing charge asymmetry (\mm{\sur>1}).  The values of \mm{n^\ast_\text{n}} increase, with increasing \mm{\sut} and  \mm{\sur}, on all centreline points (P$_1$, P$_2$, P$_3$, as shown in \fig\ref{fig:3b}).
The change in \mm{n^\ast_\text{n}} is enhanced with decreasing \mm{K}, and the maximum change is obtained at contraction point (P$_1$) as compared to other points with flow governing parameters (\mm{K, \sut, \sur}, \fig\ref{fig:3b}). It is because of the densening of the clustered excess charge with the thickening of EDL (reducing \mm{K}) and reducing flow cross-section area. Further, drastic changes in \mm{n^\ast_\text{n}} are noted at \mm{\sut=16, K=20} with increasing \mm{\sur}, as shown in \fig\ref{fig:3b} (c3), because of stronger charge attractive forces (at higher \mm{\sut}) and thin EDL (at highest \mm{K}).
For example, \mm{n^\ast_\text{n}} reduces by (54.04\%, 47.66\%, 39.61\%) and (63.3\%, 48.04\%, 44.46\%) at (P$_1$, P$_2$, P$_3$), for \mm{\sut=4} and 16 respectively for \mm{K=2}, with decreasing \mm{\sur <1} (from 1 to 0); the corresponding reduction in \mm{n^\ast_\text{n}} is noted as (49.97\%, 49.96\%, 47.08\%) and (49.6\%, 49.35\% and -404.13\%) for \mm{K=20}.
In contrast, \mm{n^\ast_\text{n}} enhances by (32.34\%, 35.05\%, 20.42\%) and (16.86\%, 21.48\% and 6.56\%) at (P$_1$, P$_2$, P$_3$), for \mm{\sut=4} and 16 respectively for \mm{K=2}, with increasing \mm{\sur >1} (from 1 to 2); the corresponding {increment} in \mm{n^\ast_\text{n}} is recorded as (49.72\%, 49.18\%,  38.84\%) and (46.10\%, 40\%, -1334.8\%) for \mm{K=20}.

\noindent 
The predictive correlation for minimum value of the excess charge (\mm{n^\ast_{\text{min}}}, \tab\ref{tab:2}) variation on the centreline (\mm{x,0}) of the asymmetrically charged microfluidic device with dimensionless governing parameters (\mm{K,\sut,\sur, d_\text{c}=0.25}) is expressed as follows.
\begin{gather}
n^\ast\smin = \alpha_2 n^\ast_{\text{min,ref}}  \label{eq:dnast}
\\
%
 \alpha_2 = \beta_1 + (\beta_2  + \beta_4 \sur) \sur + (\beta_3  + \beta_5 \gamma)\gamma + (\beta_6  + \beta_7 \sur) \sur \gamma;
\nonumber
\\ 
\beta_{\text{i}} = \sum_{{j}=1}^5 M_{\text{ij}} X^{({j}-1)};
\quad X = K^{-1}; \quad  \gamma = \sut^{-1}; \quad 1\le i\le 7 \nonumber
\\
	M = \begin{bmatrix}
		0.4067 & 1.9813 & -9.9803 &	21.614 & -16.916 \\
		0.5013 & 1.2827 & -7.8751 &	21.406 & -20.327 \\
		0.4883 & -4.1498 & 10.692 & -10.154 & 2.6274 \\
		0.0944 & -3.9881 & 24.461 & -64.573 & 59.31 \\
		-0.0481 & -30.967 & 288.04 & -958.17 & 993.26 \\
		-0.1486 & -0.8494 & -2.3027 & 40.521 & -61.006 \\
		-0.3378 & 15.439 & -104.97 & 288.63 & -270.58
	\end{bmatrix} \label{M:nast}
\end{gather}
where \mm{n^\ast_{\text{min,ref}}} is the minimum excess charge \citep[\eqn (15) in ref.][]{dhakar2022electroviscous} for liquid flow through symmetrically charged (ref: \mm{\sur=1}) contraction-expansion slit. The correlation coefficients ($M_\text{ij}$, \eqn\ref{M:nast}) are statistically obtained by performing the non-linear regression analysis using the DataFit (version 9.0, free trial) for 135 data points, with 
(\mm{\delta\smin, \delta\smax, \delta\avg,R^2}) as (-2.81\%, 2.31\%, -0.36\%, 99.98\%) between predicted (\eqn\ref{eq:dnast}) and numerical (\tab\ref{tab:2}) values. 
%
%
\subsection{Induced electric field strength ($E_{\text{x}}$)}
\label{sec:electric}
\noindent 
The transport of ions in the pressure-driven flow generates a current called streaming current (\mm{I_s}). The potential corresponding to \mm{I_s} is the streaming potential. The rate of axial variation of streaming potential is termed as induced electric field strength (\mm{E_\text{x}=-\partial U/\partial x}), which is determined from the current conservation equation (\eqn \ref{eq:7}) at the steady-state. 

\begin{figure}[h]
	\centering\includegraphics[width=1\linewidth]{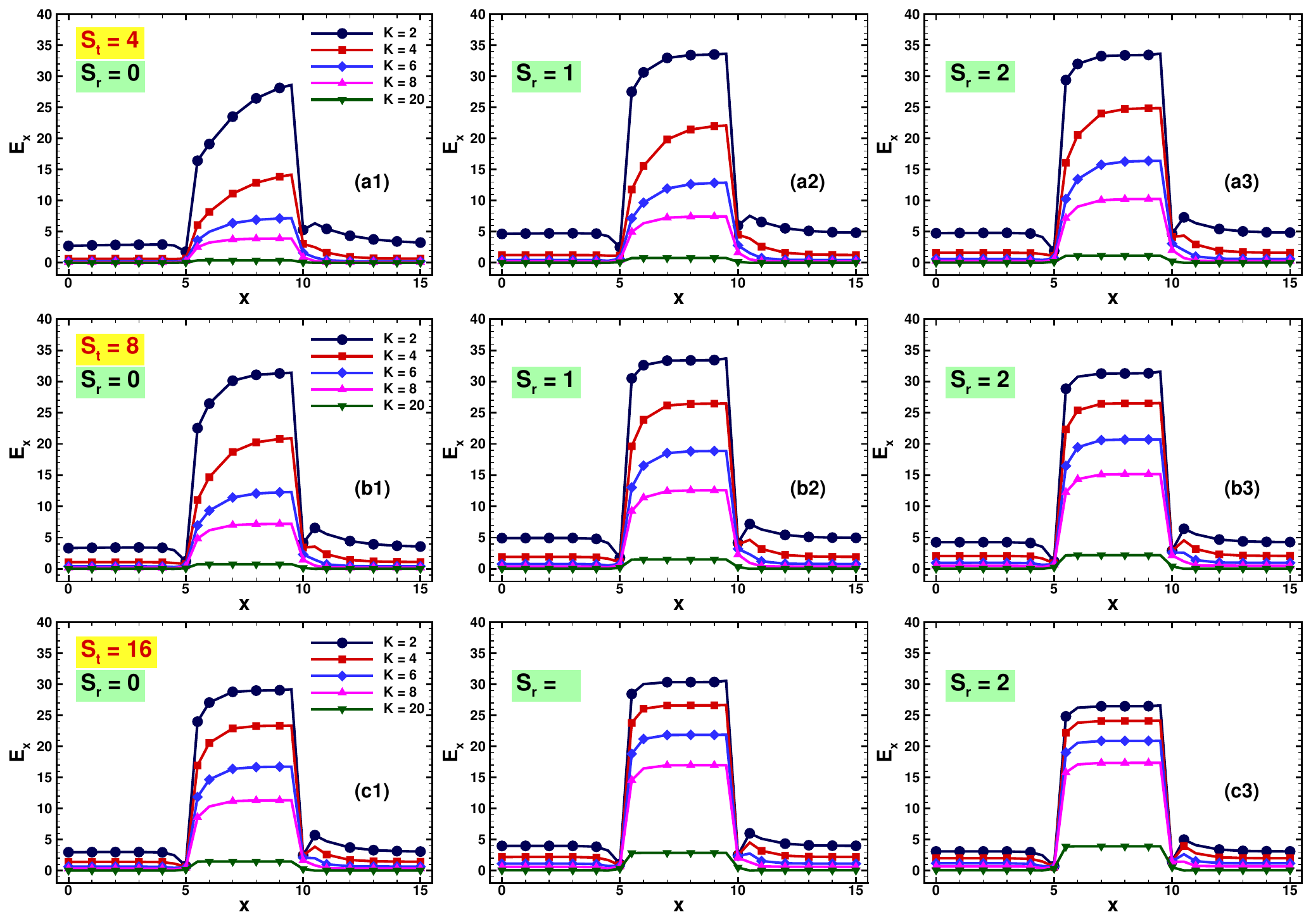}
	\caption{Dimensionless induced electric field strength ($E_{\text{x}}$) on the centreline ($x,0$) of the asymmetrically charged microfluidic device as a function of dimensionless governing parameters ($K$, $\sut$ and $\sur$).}
	\label{fig:6}
\end{figure}

\noindent 
\fig\ref{fig:6} shows the complex variation of induced electric field strength (\mm{E_\text{x}}) on the centreline (\mm{x,0}) of the asymmetrically charged microfluidic device with flow governing parameters (\mm{K, \sur, \sut}). The qualitative trends for induced electric field strength (\mm{E_\text{x}}) are the same as the literature \citep{dhakar2022electroviscous} for given ranges of governing parameters. 
In the upstream section (\mm{x\le 5}), \mm{E_\text{x}} is uniform throughout the channel, except for reducing in the last part (\mm{x\lesssim 5}). The contraction section (\mm{5\le x\le 10}) shows the complex nature of \mm{E_\text{x}} along the length, such as a steep increase near the sudden contraction, followed by a smooth increase with a reducing gradient in the middle part and a sudden drop at the end of the contraction region. Further, in the downstream section (\mm{10\le x\le 15}), \mm{E_\text{x}} increases initially near the expansion section, then decreases slowly and becomes constant in the second half part of the downstream region (as shown in \fig\ref{fig:6}).
The change in induced electric field strength (\mm{E_\text{x}}) is maximum at \mm{K=20} over the ranges of conditions.
In general, \mm{E_\text{x}} is maximum in the contraction, as compared to up/downstream sections (\fig\ref{fig:6}), because of enhanced convection with a reduction in the cross-section of flow area.  
\begin{table}
\centering\caption{Maximum induced electric field strength (\mm{E_{\text{x,max}}}) on the centreline (\mm{x,0}) of the asymmetrically charged microfluidic device (underlined data shows highest \mm{E_{\text{x,max}}} for given \mm{K} and \mm{\sut}).}\label{tab:emax}
\scalebox{0.85}
{
\begin{tabular}{|r|r|r|r|r|r|r|r|r|r|r|}
		\hline
		$\sut$	&	$K$	&	\multicolumn{9}{c|}{$E_{\text{x,max}}$}	\\\cline{3-11}
		&		&	$\sur=0$	&	$\sur=0.25$	& $\sur=0.50$ &	$\sur=0.75$	& $\sur=1$ & $\sur=1.25$ & $\sur=1.50$ & $\sur=1.75$ & $\sur=2$  \\\hline
		0	&	$\infty$	&	0	& 0		& 0		& 0		& 0 	& 0  &  0  &  0   &  0	\\\hline
		4	& 2	& 28.6420	& 31.1030	& 32.4500	& 33.2330	& 33.6690	& 33.8710	& \htxt{33.9110}	& 33.8370	& 33.6810 \\
		& 4	& 14.1750	& 16.8620	& 19.0500	& 20.7740	& 22.1050	& 23.1220	& 23.8960	& 24.4780	& \htxt{24.9110} \\
		& 6	& 7.1829	& 8.8130	& 10.3170	& 11.6800	& 12.8980	& 13.9730	& 14.9120	& 15.7240	& \htxt{16.4230} \\
		& 8	& 3.9032	& 4.8487	& 5.7627	& 6.6374	& 7.4672	& 8.2472	& 8.9755	& 9.6499	& \htxt{10.2710} \\
		& 20	& 0.3821	& 0.4777	& 0.5731	& 0.66825	& 0.7630	& 0.8573	& 0.9509	& 1.0439 	& \htxt{1.1362} \\\hline
		8	& 2	& 31.4090	& 32.9170	& 33.6810	& \htxt{33.8600}	& 33.6930	& 33.3190	& 32.8160	& 32.2270	& 31.5850 \\
		& 4	& 20.9100	& 23.3350	& 24.9110	& 25.9000	& 26.4760	& 26.7570	& \htxt{26.8280}	& 26.7460	& 26.5520 \\
		& 6	& 12.3020	& 14.5960	& 16.4230	& 17.8270	& 18.8750	& 19.6350	& 20.1640	& 20.5120 	& \htxt{20.7170} \\
		& 8	& 7.2044	& 8.8252	& 10.2710	& 11.5250	& 12.5840	& 13.4610	& 14.1730	& 14.7400	& \htxt{15.1820} \\
		& 20	& 0.7576	& 0.9476	& 1.1362	& 1.3225	& 1.5060	& 1.6858	& 1.8612	& 2.0319 	& \htxt{2.1974} \\\hline
		16	& 2	& 29.2120	& 31.1100	& \htxt{31.5850}	& 31.2770	& 30.5800	& 29.6770	& 28.6730	& 27.6340	& 26.6000 \\
		& 4	& 23.3610	& 25.5150	& 26.5520	& \htxt{26.8470}	& 26.6800	& 26.2320	& 25.6190	& 24.9140 	& 24.1650 \\
		& 6	& 16.7450	& 19.1930	& 20.7170	& 21.5420	& 21.8790	& \htxt{21.8890}	& 21.6850	& 21.3430 	& 20.9140 \\
		& 8	& 11.3330	& 13.5460	& 15.1820	& 16.2960	& 16.9880	& 17.3650	& \htxt{17.5120}	& 17.4960 	& 17.3660 \\
		& 20	& 1.4662	& 1.8371	& 2.1974	& 2.5416	& 2.8655	& 3.1659	& 3.4412	& 3.6908 	& \htxt{3.9150} \\\hline
\end{tabular}
}
\end{table}
\\\noindent 
The maximum values of induced electric field strength (\mm{E_\text{x,max}}) on the centreline ($x,0$) of the asymmetrically charged device listed in \tab\ref{tab:emax} show the complex dependence on the broader ranges of dimensionless parameters (\mm{K, \sur, \sut}). The \mm{E_\text{x,max}} increases with decreasing $K$ (thickening of EDL) and with increasing \mm{\sur} and \mm{\sut}, irrespective of other parameters. It is because of enhanced surface charge density (\mm{\sur}, \mm{\sut}) which increases the available excess charge ((\mm{n^\ast}), as shown in \fig\ref{fig:5} and \tab\ref{tab:2}) for transport that strengthens the induced electric field strength (\mm{E_\text{x}}). However, \mm{E_\text{x}} decreases at higher charge density (\mm{\sur}, \mm{\sut}) due to stronger electrostatic interactions close to the walls, which retards the flow of excess charge (as shown in \fig\ref{fig:6}).
For instance, \mm{E_\text{x,max}} reduces by (14.93\%, 6.78\%, 4.47\%) and (49.92\%, 49.69\%, 48.83\%) at \mm{K=2} and 20, respectively, for (\mm{\sut=4, 8, 16}) with decreasing charge asymmetry (\mm{\sur}) from 1 to 0. The corresponding changes with increasing charge asymmetry (\mm{\sur>1}) from 1 to 2 are recorded as (0.04\%, -6.26\%, -13.02\%) and (48.91\%, 45.91\%, 36.63\%) at \mm{K=2} and 20, respectively. Further, \mm{E_\text{x,max}} varies by (17.59\%, 0.56\%, -8.94\%) and (197.36\%, 190.04\%, 167.02\%) at \mm{K=2} and 20, respectively, for (\mm{\sut=4, 8, 16}) with overall increase in the charge asymmetry (\mm{\sur}) from 0 to 2. 
It is also noted that the change in \mm{E_\text{x,max}} is greater at \mm{\sut=4}  (\tab\ref{tab:emax}), as compared to \mm{\sut>4} irrespective of the other conditions (\mm{K, \sur}).
%
%

\noindent
Subsequently, \fig\ref{fig:13b} maps the conditions (\mm{K, \sut, \sur}) which yield the largest value of induced electric field strength (\mm{E_\text{x,max,h}}). The largest \mm{E_\text{x,max}} is obtained at \mm{\sut=8}, as compared to \mm{\sut=4, 16} at \mm{\sur=1} which suggest initiation of overlapping of EDL from \mm{\sut\approx 8} at \mm{\sur=1}.
However, the largest \mm{E_\text{x,max}} is obtained with increasing \mm{\sut}, except at \mm{K=20}, for \mm{\sur<2} (as shown in \tab\ref{tab:emax}). The shift in \mm{E_\text{x,max,h}} with \mm{\sur} is greater at the lower \mm{K}, and that is smaller at the higher \mm{K}, as shown in \fig\ref{fig:13b}, because \mm{K} is inversely proportional to the EDL thickness. For instance, \mm{E_\text{x,max,h}} is noted as 33.911 (at \mm{\sur=1.50}), 33.860 (at \mm{\sur=0.75}), and 31.585 (at \mm{\sur=0.50}) for \mm{\sut=4}, 8, and 16, respectively, at lower \mm{K=2}. Similarly, \mm{E_\text{x,max,h}} is noted as 10.271 (at \mm{\sur=2}), 15.182 (at \mm{\sur=2}), and 17.512 (at \mm{\sur=1.50}) for \mm{\sut=4}, 8, and 16, respectively, at higher \mm{K=8}. Thus, it is found that, even at the higher \mm{K=8}, EDL overlaps at higher top wall surface charge density (\mm{\sut=16}) and surface charge ratio (\mm{\sur=1.50}) (as shown in \fig\ref{fig:13} (b)). 
It is because strengthening in the charge attractive forces retards the flow of excess charge in the EDL at higher \mm{\sut} and \mm{\sur} (as shown in \fig\ref{fig:6} and \tab\ref{tab:emax}). 
\begin{figure}[t]
	\centering\includegraphics[width=1\linewidth]{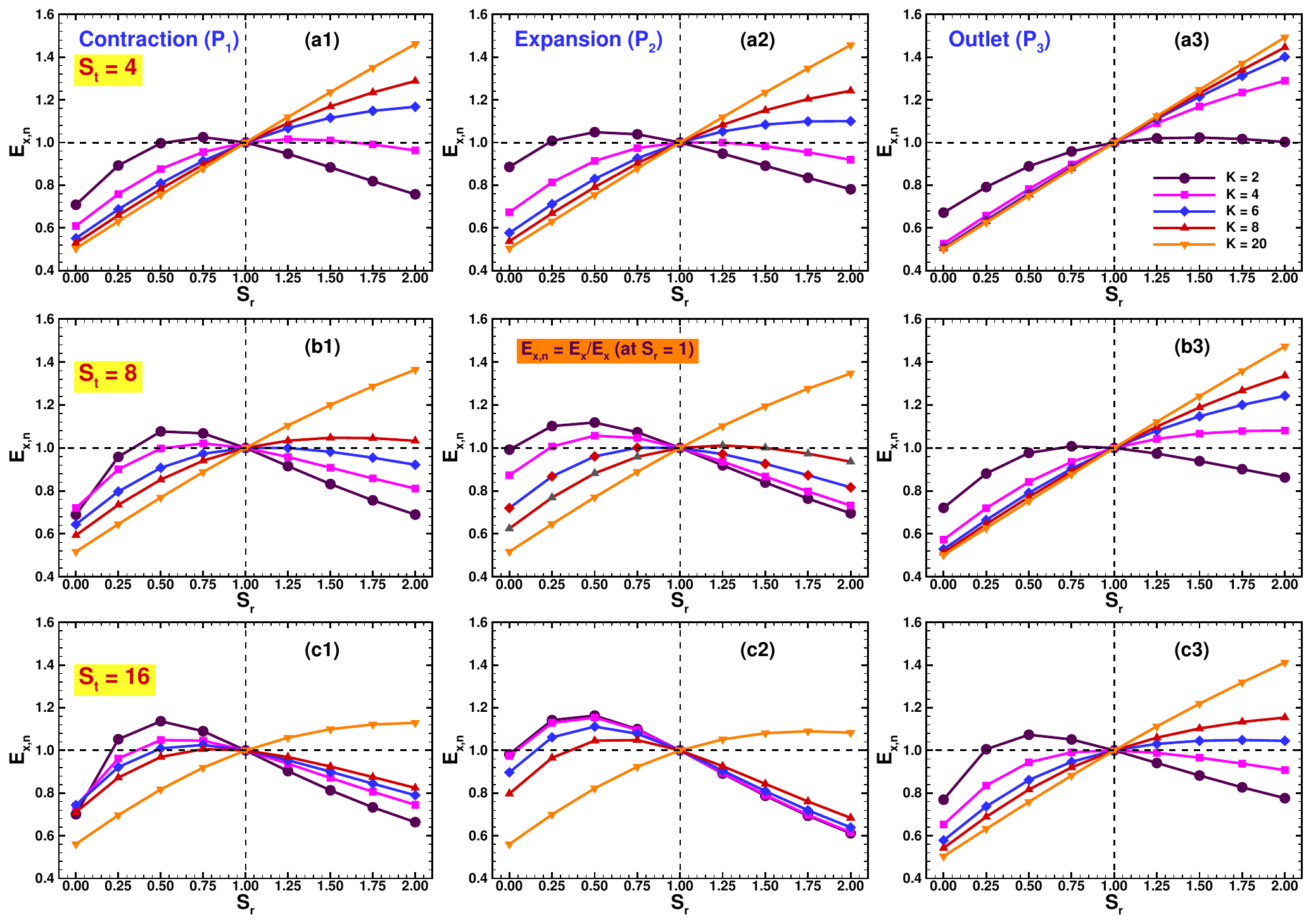}
	\caption{Normalized  induced electric field strength (\mm{E_\text{x,n}}, \eqn\ref{eq:Phin}) variation on contraction ($P_1$, 1st column), expansion ($P_2$, 2nd column), and outlet ($P_3$, 3rd column) centreline points (refer \fig\ref{fig:1}) of the microfluidic device with \mm{\sur} at \mm{2\le K\le 20}  and \mm{4\le \sut\le 16}.}
	\label{fig:3c}
\end{figure} 
\\\noindent
Further, \fig\ref{fig:3c} depicts the normalized  induced electric field strength (\mm{E_\text{x,n}}, \eqn\ref{eq:Phin}) variation on the contraction (P$_1$), expansion (P$_2$), and outlet (P$_3$) points on the centreline (refer \fig\ref{fig:1}) of the microfluidic device with flow governing parameters (\mm{K, \sut, \sur}). In general, \mm{E_\text{x,n}} decreases (for \mm{\sur<1}) and increases (for \mm{\sur>1}) with increasing \mm{K} for a given \mm{\sut}. Further, \mm{E_\text{x,n}} increases with increasing \mm{\sut} and \mm{\sur} for all points (P$_1$, P$_2$, P$_3$), as shown in \fig\ref{fig:3c}. 
For instance, \mm{E_\text{x,n}} reduces by (29.15\%, 11.45\%, 32.86\%) and (29.92\%, 1.85\%, 23.09\%) at (P$_1$, P$_2$, P$_3$) with decreasing charge asymmetry (\mm{\sur=1} to 0) for \mm{\sut=4} and 16, and \mm{K=2}; the corresponding reduction in \mm{E_\text{x,n}} at \mm{K=20} are noted by (49.57\%, 49.57\%, 49.98\%) and (43.90\%, 43.94\%, 49.63\%). On the other hand, \mm{E_\text{x,n}} changes by (-24.24\%, -21.90\%, 0.25\%) and (-33.61\%, -38.84\%, -22.4\%) with increasing charge asymmetry (\mm{\sur=1} to 2) at (P$_1$, P$_2$, P$_3$) for \mm{\sut=4} and 16, and \mm{K=2}; the corresponding increase in \mm{E_\text{x,n}} at \mm{K=20} are noted by (46.15\%, 45.65\%, 49.28\%) and (12.92\%, 8.28\%, 41.18\%). The change in \mm{E_\text{x,n}} is smaller at low \mm{K}. The maximum change in \mm{E_\text{x,n}} is obtained at P$_{3}$ (\fig\ref{fig:3c}), as compared to P$_{1}$ and P$_{2}$, with the flow governing parameters (\mm{K, \sut, \sur}) because of the reduction in the cross-section area increases the convective velocity.

\noindent 
The predictive correlation for maximum induced electrical field strength (\mm{E_\text{x,max}}, \tab\ref{tab:emax}) variation on the centreline (\mm{x,0}) of the asymmetrically charged contraction-expansion microfluidic device with dimensionless governing parameters (\mm{K,\sut,\sur, d_\text{c}=0.25}) is expressed as follows.
\begin{gather}
	E_{\text{x,max}} = \alpha_3 E_\text{x,max,ref}   \label{eq:emax}
	\\
	\alpha_3 = \beta_1 +  (\beta_2 + \beta_4 \sur)\sur + (\beta_3 + \beta_5 \gamma)\gamma + (\beta_6 +\beta_7 \sur \gamma^{-3} )\sur \gamma
\nonumber
	\\
	\beta_{\text{i}} = \sum_{{j}=1}^5 M_{\text{ij}} X^{({j}-1)};\quad X = K^{-1}; \quad  \gamma = \sut^{-1}; \quad 1\le i\le 7 \nonumber
\\
	M = \begin{bmatrix}
	0.2776 & 6.0816 & -11.591 & -17.295 & 44.379 \\
	0.3448 & 7.0632 & -80.208 & 263.14 & -266.39 \\
	4.0066 & -130.01 & 836.81 & -2046 & 1708 \\
	0.0191 & -0.3723 & -6.2271 & 32.957 & -39.382 \\
	-11.359 & 384.96 & -2841.2 & 7728.8 & -6980.7 \\
	0.1174 & -12.11 & 224.13 & -882.59 & 971.89 \\
	0.000621 & -0.0238 & 0.19521 & -0.5872 & 0.573
	\end{bmatrix} \label{M:emax}
\end{gather}
where \mm{E_\text{x,max,ref}} is the maximum induced electric field strength \citep[\eqn (16) in ref.][]{dhakar2022electroviscous} for liquid flow through symmetrically charged (ref: \mm{\sur=1}) contraction-expansion slit.
The correlation coefficients ($M_\text{ij}$, \eqn\ref{M:emax}) are statistically obtained by performing the non-linear regression analysis using the DataFit (version 9.0, free trial) for 135 data points, with 
(\mm{\delta\smin, \delta\smax, \delta\avg,R^2}) as (-4.87\%, 4.73\%, -0.54\%, 99.94\%) between  predicted (\eqn\ref{eq:emax}) and numerical (\tab\ref{tab:emax}) values. 
%
%
\\\noindent
In general, the previous sections have elucidated the complex dependence of electrostatic and ionic fields (electrical potential, excess charge, and induced electrical field strength) on the dimensionless flow governing parameters (\mm{K, \sut, \sur}).  The subsequent section explores the corresponding influences on the hydrodynamic field.
%
\subsection{Pressure ($P$) distribution}
\label{sec:pressure}
\begin{figure}[tb]
	\centering\includegraphics[width=1\linewidth]{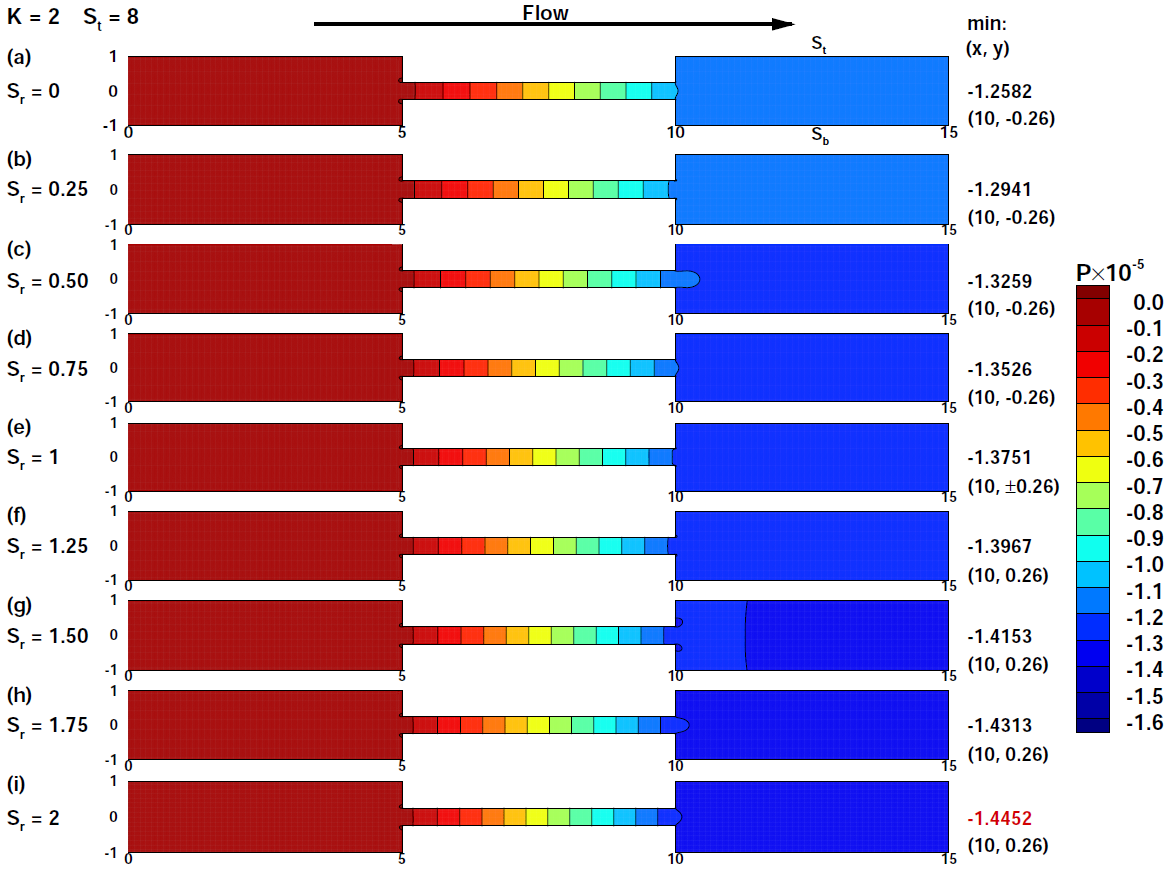}
	\caption{Dimensionless pressure (\mm{P\times 10^{-5}}) distribution as a function of \mm{\sur} at \mm{\sut=8} and \mm{K=2}.}
	\label{fig:7}
\end{figure}

%
\noindent \fig\ref{fig:7} shows the pressure ($P$) distribution in an asymmetrically charged microfluidic device for \mm{0\le \sur\le 2} at \mm{K=2} and \mm{\sut=8}; the contour profiles are qualitatively similar for other flow governing parameters (\mm{2\le K\le 20}, \mm{4\le \sut\le 16}, \mm{0\le \sur\le 2}) and thus not presented here. 
The pressure profiles have shown consistent features with the literature \citep{dhakar2022electroviscous} for limiting condition (\mm{2\le K\le 20}, \mm{4\le \sut\le 16}, \mm{\sur=1}). In general, as expected, the pressure has reduced over the length of the device, and the pressure gradient is maximum in the contraction than the upstream or downstream section (\fig\ref{fig:7}).
In addition to the reducing flow cross-section area in the contraction section, the additional resistance imposed by the streaming potential reduces the pressure, as discussed earlier in the section \ref{sec:potential} (as shown in \fig\ref{fig:3}).
Further, the pressure magnitude (\mm{|P|}) increases with increasing surface charge density ratio (\mm{\sur}) because of enhancement in the charge attractive forces near the walls that increases the wall stress and imposes an additional resistance on the flow. 
For instance, the pressure drop (\mm{|\Delta P|\times 10^{-5}}) increases from 1.2582 to 1.4452 with increasing \mm{\sur} from 0 to 2 at fixed \mm{K} and \mm{\sut} (as shown in \fig\ref{fig:7}). Further, the maximum pressure drop 
(\mm{|\Delta P|_\text{max} \times 10^{-5}=1.4452}) is obtained at a position (10, 0.26) near the expansion section of the microfluidic device at \mm{K=2}, \mm{\sut=8}, and \mm{\sur=2} (as shown in \fig\ref{fig:7}(i)) for the ranges of conditions explored herein. {Furthermore, the  maximum pressure drop (\mm{|\Delta P|_\text{max}\times 10^{-5}=1.5702}) in the microfluidic device over the whole range of conditions is obtained at \mm{K=2}, \mm{\sut=16} and \mm{\sur=2}.} 
%
%
%
%
\begin{figure}[h]
	\centering\includegraphics[width=1\linewidth]{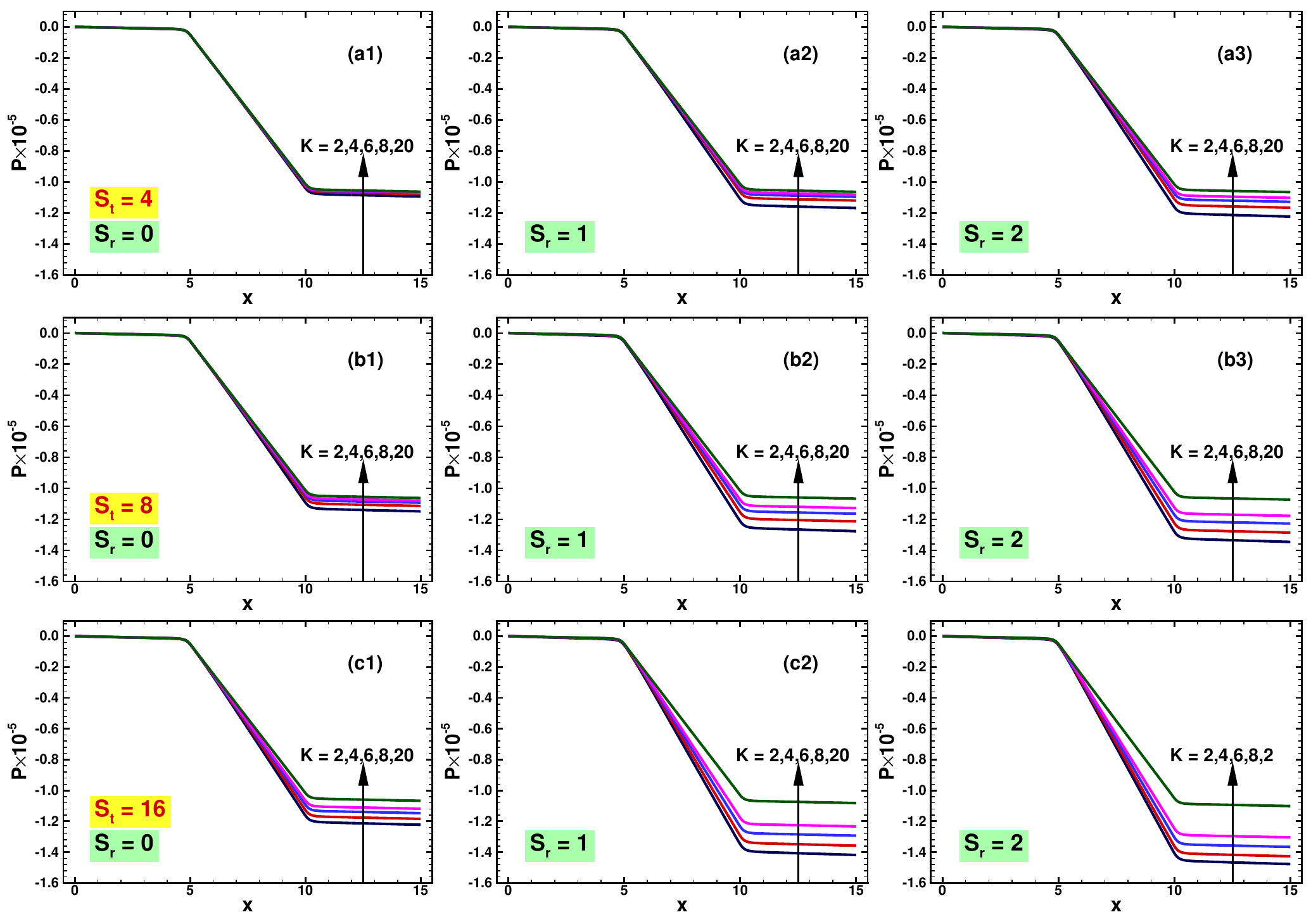}
	\caption{Pressure (\mm{P}) variation on the centreline (\mm{0\le x\le L,0}) of the asymmetrically charged microfluidic device as a function of governing parameters (\mm{K, \sut, \sur}).}
	\label{fig:8}
\end{figure} 

\noindent
Subsequently, \fig\ref{fig:8} depicts the pressure (\mm{P}) variation on the centreline (\mm{0\le x\le L,0}) of the asymmetrically charged microfluidic device for broader conditions (\mm{2\le K\le 20; 4\le \sut\le 16; 0\le \sur\le 2}). The magnitude of the pressure (\mm{|P|}) increases along the length (\mm{0\le x\le L}) of the microfluidic device (\fig\ref{fig:8}), irrespective of the flow conditions. It is because enhancement in the resistance imposed by the streaming potential increases along the length (as shown in \fig\ref{fig:3}), in addition to hydrodynamic resistance. Further, both resistances impede the flow with decreasing flow area. As a result, the pressure gradient (\mm{|dP/dx|}) is sharp in contraction compared to up/downstream sections.
The pressure variations strongly depend on the governing parameters (\mm{K, \sut, \sur}). The pressure magnitude increases with decreasing \mm{K} (inverse Debye length) due to enhanced streaming potential with the thickening of EDL. Further, \mm{|P|} increases with increasing \mm{\sur} and \mm{\sut} because strengthening the attractive charge forces near the walls increases the wall stress and imposes an additional resistance to the pressure-driven flow (\fig\ref{fig:8}). The qualitative trends of pressure variation with the flow parameters are consistent with the literature \citep{davidson2007electroviscous,dhakar2022electroviscous}.
%
%
%
\\
\noindent 
\tab\ref{tab:4} presents the pressure drop (\mm{|\Delta P| \times 10^{-5}}) over the length (\mm{x,0}) of an asymmetrically charged microfluidic device for the range of parameters (\mm{K, \sut, \sur}) governing the electroviscous and non-electroviscous (\mm{\sut=0} or \mm{K=\infty}) flows. The pressure drop (\mm{|\Delta P|}) decreases with increasing \mm{K} and achieving the value that for non-electroviscous condition (\mm{K\rightarrow\infty}). For instance, the pressure drop (\mm{|\Delta P|}) decreases by (8.94\%, 16.43\%, 23.73\%) for \mm{(\sut=4, 8, 16)} with increasing \mm{K} from 2 to 20 at symmetric charged (\mm{\sur=1}) condition \citep{dhakar2022electroviscous}. Further, \mm{|\Delta P|} shows minimum reduction with the charge asymmetry (\mm{\sur\neq 1}) at \mm{K=20}. For example, \mm{|\Delta P|} decreases by (6.35\%, 10\%, 13.8\%) and (0.09\%, 0.37\%, 1.36\%) for \mm{(\sut=4, 8, 16)} at \mm{K=2} and 20 with decreasing \mm{\sur} from 1 to 0.  On the other hand, \mm{|\Delta P|} enhances by (4.71\%, 5.4\%, 4.17\%) and (0.16\%, 0.58\%, 1.79\%) for \mm{(\sut=4, 8, 16)} at \mm{K=2} and 20 with increasing \mm{\sur} from 1 to 2. Overall increasing \mm{\sur} from 0 to 2 enhances \mm{|\Delta P|} by (11.81\%, 17.12\%, 20.85\%) and (0.25\%, 0.96\%, 3.2\%) for \mm{(\sut=4, 8, 16)} at \mm{K=2} and 20. It can thus be noted that \mm{|\Delta P|} changes significantly at \mm{\sut=16} as compared to \mm{\sut=4} and 8, irrespective of other conditions (\mm{K, \sur}); and \mm{|\Delta P|} increases with increasing \mm{\sut}, as stronger electrostatic interactions at higher surface charge density provide additional resistance to the pressure-driven flow. 
%
%
\begin{table}
	\centering\caption{Asymmetric charge effects (\mm{0\le \sur\le 2}) on the pressure drop (\mm{10^{-5}|\Delta P|}) on the centreline (\mm{x,0}) of the microfluidic device in electroviscous (\mm{2\le K\le 20; 4\le \sut\le 16}) and non-electroviscous (\mm{\sut=0} or \mm{K=\infty}) flows (underlined data shows the maximum value for given \mm{K} and \mm{\sut}).}\label{tab:4}
	\scalebox{0.85}
	{
		\begin{tabular}{|r|r|r|r|r|r|r|r|r|r|r|}
			\hline
			$\sut$	&	$K$	&	\multicolumn{9}{c|}{$10^{-5}|\Delta P|$}	\\\cline{3-11}
			&		&	$\sur=0$	&	$\sur=0.25$	& $\sur=0.50$ &	$\sur=0.75$	& $\sur=1$ & $\sur=1.25$ & $\sur=1.50$ & $\sur=1.75$ & $\sur=2$  \\\hline
			0	&	$\infty$	&	1.0616 	&  1.0616  &  1.0616  &  1.0616  &  1.0616  & 1.0616      &   1.0616     &  1.0616       &  1.0616    	  \\\hline
			4	& 2	& 1.0931	& 1.1120	& 1.1315	& 1.1501	& 1.1672	& 1.1828	& 1.1971	& 1.2102	& \htxt{1.2222} \\ 
			& 4	& 1.0774	& 1.0861	& 1.0962	& 1.1073	& 1.1189	& 1.1307	& 1.1425	& 1.1540	& \htxt{1.1652} \\
			& 6	& 1.0706	& 1.0756	& 1.0814	& 1.0880	& 1.0952	& 1.1029	& 1.1108	& 1.1190	& \htxt{1.1272} \\
			& 8	& 1.0667	& 1.0695	& 1.0728	& 1.0767	& 1.0811	& 1.0858	& 1.0909	& 1.0962	& \htxt{1.1018} \\
			& 20	& 1.0619	& 1.0621	& 1.0623	& 1.0626	& 1.0629	& 1.0633	& 1.0637	& 1.0641	& \htxt{1.0646} \\\hline
			8	& 2	& 1.1489	& 1.1876	& 1.2222	& 1.2516	& 1.2766	& 1.2980	& 1.3164	& 1.3321	& \htxt{1.3456} \\
			& 4	& 1.1140	& 1.1393	& 1.1652	& 1.1903	& 1.2136	& 1.2348	& 1.2539	& 1.2708	& \htxt{1.2857} \\
			& 6	& 1.0931	& 1.1094	& 1.1272	& 1.1457	& 1.1640	& 1.1816	& 1.1981	& 1.2135	& \htxt{1.2275} \\
			& 8	& 1.0802	& 1.0902	& 1.1018	& 1.1143	& 1.1274	& 1.1405	& 1.1535	& 1.1660	& \htxt{1.1779} \\
			& 20	& 1.0629	& 1.0637	& 1.0646	& 1.0657	& 1.0669	& 1.0683	& 1.0698	& 1.0714	& \htxt{1.0731} \\\hline
			16	& 2	& 1.2222	& 1.2910	& 1.3456	& 1.3869	& 1.4179	& 1.4407	& 1.4573	& 1.4690	& \htxt{1.4770} \\
			& 4	& 1.1839	& 1.2376	& 1.2857	& 1.3257	& 1.3577	& 1.3825	& 1.4016	& 1.4159	& \htxt{1.4266} \\
			& 6	& 1.1470	& 1.1879	& 1.2275	& 1.2628	& 1.2928	& 1.3174	& 1.3373	& 1.3532	& \htxt{1.3657} \\
			& 8	& 1.1181	& 1.1475	& 1.1779	& 1.2069	& 1.2329	& 1.2555	& 1.2747	& 1.2908	& \htxt{1.3042} \\
			& 20	& 1.0668	& 1.0697	& 1.0731	& 1.0771	& 1.0815	& 1.0862	& 1.091	& 1.0960	& \htxt{1.1009} \\\hline
		\end{tabular}
	}
\end{table}
\begin{figure}[h]
	\centering\includegraphics[width=1\linewidth]{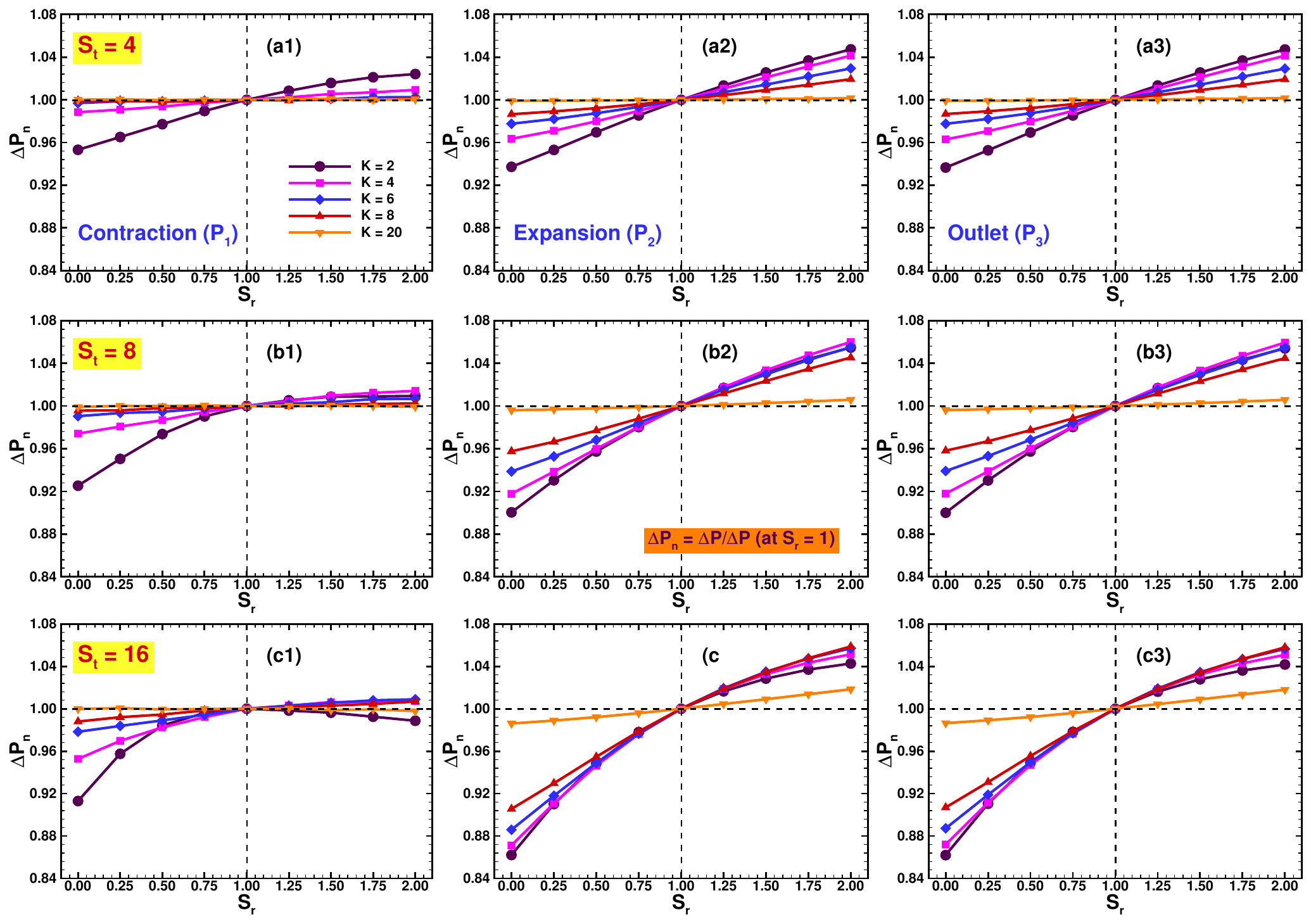}
	\caption{Normalized pressure drop (\mm{\Delta P_\text{n}},) variation on contraction ($P_1$, 1st column), expansion ($P_2$, 2nd column), and outlet ($P_3$, 3rd column) centreline locations of the microfluidic device with \mm{\sur} at \mm{2\le K\le 20} and \mm{4\le \sut\le 16}.}
	\label{fig:8a}
\end{figure} 

\noindent
Further, \fig\ref{fig:8a} depicts the normalized  pressure drop (\mm{\Delta P_\text{n}}, \eqn\ref{eq:Phin}) variation on the contraction (P$_1$), expansion (P$_2$), and outlet (P$_3$) points on the centreline (refer \fig\ref{fig:1}) of the microfluidic device with flow governing parameters (\mm{K, \sut, \sur}). The normalized pressure drop (\mm{\Delta P_\text{n}}) increases with increasing \mm{K} (or EDL thinning) for lower charge asymmetry (\mm{\sur<1}) but decreases with increasing \mm{K} for higher charge asymmetry (\mm{\sur>1}). The normalized pressure drop (\mm{\Delta P_\text{x,n}}) increases with increasing of both \mm{\sut} and \mm{\sur} at all points (P$_1$, P$_2$, P$_3$). 
For instance, \mm{\Delta P_\text{x,n}} reduces by (4.68\%, 6.28\%, 6.35\%) and (8.7\%, 13.79\%, 13.8\%) at \mm{\sut=4} and 16 on the points (P$_1$, P$_2$, P$_3$) at \mm{K=2} with decreasing charge asymmetry from 1 to 0 (\mm{\sur<1}); the corresponding changes in \mm{\Delta P_\text{x,n}} at \mm{K=20} are (-0.01\%, 0.10\%, 0.09\%) and (0.02\%, 1.4\%, 1.36\%) at \mm{\sut=4} and 16. On the other hand, \mm{\Delta P_\text{x,n}} changes by (2.41\%, 4.73\%, 4.71\%) and (-1.14\%, 4.26\%, 4.17\%) at \mm{\sut=4} and 16 on the points (P$_1$, P$_2$, P$_3$) at \mm{K=2} with increasing charge asymmetry from 1 to 2 (\mm{\sur>1}); the corresponding changes in \mm{\Delta P_\text{x,n}} at \mm{K=20} are (0.05\%, 0.16\%, 0.16\%) and (-0.22\%, 1.83\%, 1.79\%) at \mm{\sut=4} and 16.
Thus, the change in the pressure drop increases with decreasing \mm{K}  (\fig\ref{fig:8a}) because of the thickening of EDL. {Further, the maximum change in the pressure drop is obtained at P$_2$ as compared to P$_1$; on the other hand, negligible change is obtained from P$_2$ to P$_3$ with flow governing parameters (\mm{K, \sut, \sur}) (\fig\ref{fig:8a}).} 
{It is because reducing flow area and clustering of excess charge enhance both hydrodynamic and charge attractive resistance at P$_2$ than P$_1$ and P$_3$ which retards the flow of electrolyte liquid in the device and increases pressure drop.}
\\\noindent
The above discussion has shown that the total electrical potential, excess charge, induced electric field strength, and pressure fields strongly depend on the flow governing parameters (\mm{K, \sut, \sur}).
%
\subsection{Electroviscous correction factor ($Y$)}
\label{sec:ECF}
\noindent 
In the electroviscous flow (EVF), the streaming potential gradient (or induced electric field strength) imposes additional hydrodynamic resistance due to the convective transport of the excess charged ions in the pressure-driven flow through the microfluidic device. This extra resistance manifests the pressure drop (\mm{\Delta P}) for EVF (i.e., \mm{\sut>0}), which is higher than that of the pressure drop (\mm{\Delta P_0}) for non-EVF (i.e., \mm{\sut=0}, or \mm{K\rightarrow \infty}) at the fixed volumetric flow rate (\mm{Q}). In general, the electroviscous effect (EVE) is quantified \citep{davidson2007electroviscous,davidson2008electroviscous,bharti2008steady,dhakar2022electroviscous} in terms of the apparent viscosity (\mm{\mu _\text{eff}}) yielding the pressure drop (\mm{\Delta P}) under non-EVF (i.e., \mm{\sut=0}).
The non-linear advection term in the momentum transport equation (\eqn\ref{eq:4}) becomes negligible in the steady laminar pressure-driven microfluidic flow at a low $Re$. In turn, the relative enhancement in the viscosity (\mm{\mu_{\text{eff}}/\mu}) of fluid relates to the relative enhancement in the pressure drop (\mm{\Delta P/\Delta P_0}) at low \mm{Re}, under otherwise identical conditions.  
Thus, the \textit{electroviscous correction factor} (EVCF, \mm{Y}) is defined as follows.
\begin{gather}
	Y=\frac{\mu_{\text{eff}}}{\mu}=\frac{\Delta P}{\Delta P_{\text{0}}}
	\label{eq:27}
\end{gather}
where, \mm{\Delta P} and \mm{\Delta P_0} are the pressure drop for EVF  (\mm{\sut>0}) and non-EVF (\mm{\sut=0} or \mm{K=\infty}), respectively. 
%
%
Further, \mm{\mu} and \mm{\mu_{\text{eff}}} represent the liquid viscosities yielding the pressure drop \mm{\Delta P_0} and \mm{\Delta P~(=\Delta P_{\text{eff}})} both in the absence of electric field (non-EVF, \mm{E_\text{x}=0}).
\begin{figure}[h]
	\centering
	{\includegraphics[width=1\linewidth]{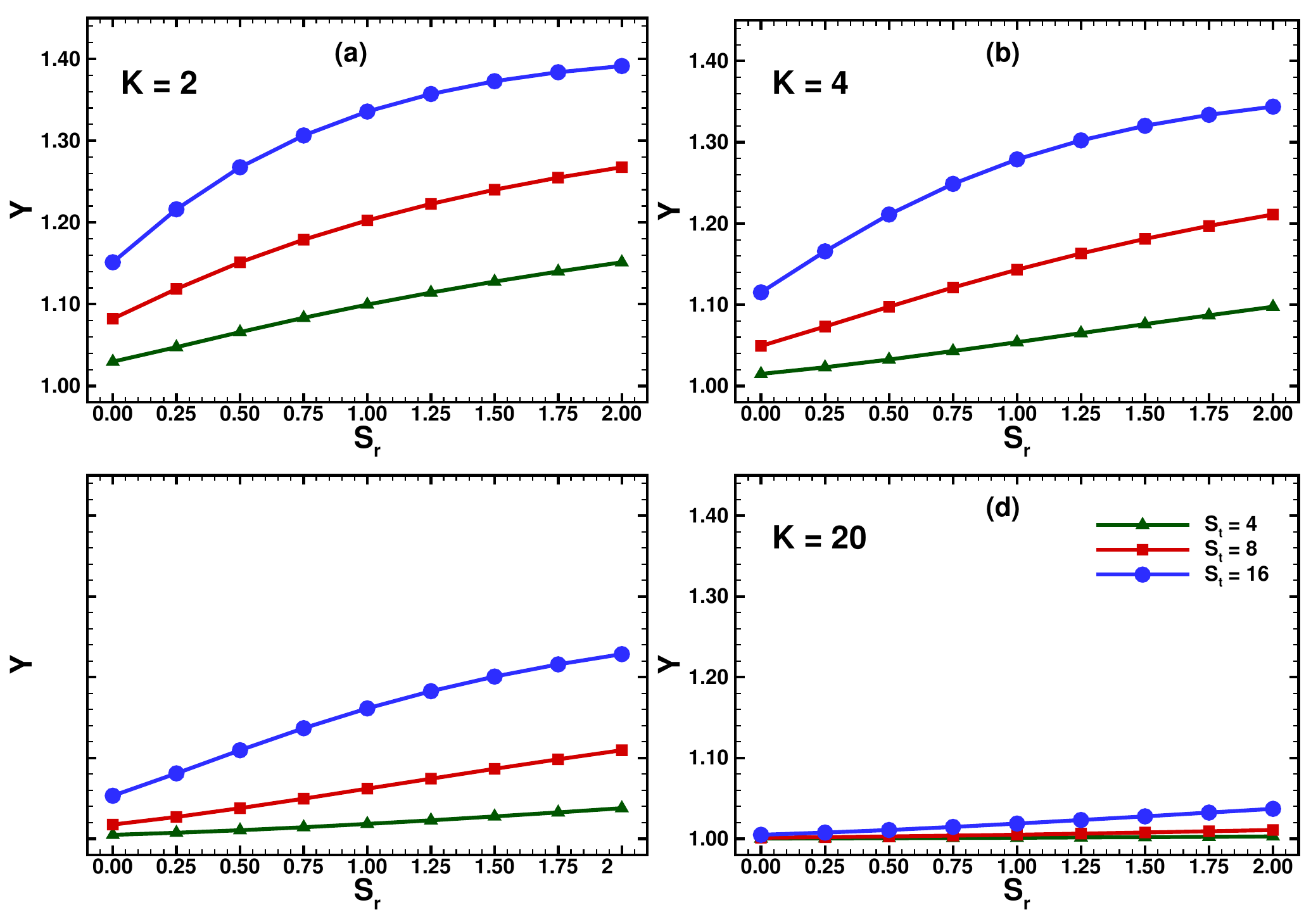}}
	\caption{Electroviscous correction factor (\mm{Y}) as a function of flow governing parameters (\mm{K, \sut, \sur}).}
	\label{fig:11}
\end{figure} 

\noindent
\fig\ref{fig:11} shows the electroviscous correction factor (EVCF, $Y$) as a function of the flow governing parameters (\mm{2\le K\le 20}, \mm{4\le \sut\le 16}, \mm{0\le \sur\le 2}). In general, the electroviscous correction factor (EVCF, $Y$, \eqn\ref{eq:27})  enhances with increasing \mm{\Delta P > \Delta P_0} as \mm{\Delta P_0} remains unchanged for \mm{\sut = 0} under otherwise identical conditions.
The factor (\mm{Y}) depicts the complex non-linear dependency on the governing parameters (\mm{K,\sut,\sur}). For example, \mm{Y} varies inversely proportional with increasing \mm{K} (thinning of EDL), i.e., \mm{Y\rightarrow 1} as $K\rightarrow \infty$ (non-EVF), as the pressure drop is found to decrease with increasing \mm{K} (refer \fig\ref{fig:8} and \tab\ref{tab:4}). Further, \mm{Y} depicts proportional dependency on \mm{\sut} and \mm{\sur}, i.e., \mm{Y\rightarrow 1}, as \mm{\sut\rightarrow 0} because of strengthening in the charge attractive forces near the walls with increasing charge density (\mm{\sut}, \mm{\sur}) enhances the wall stress, and hence, pressure drop (refer \fig\ref{fig:11}). 
{Over the range of conditions explored in this work, the electroviscous correction factor (EVCF, \mm{Y}) enhances maximally by 20.85\% (at \mm{K=2}, \mm{\sut=16}) and 34.16\% (at \mm{\sut=16}, \mm{\sur=2}) with the overall charge asymmetry (\mm{0\le \sur\le 2}), and with \mm{K} (from 20 to 2), respectively. Further, the maximum increment in \mm{Y} is noted as 39.13\% at \mm{K=2}, \mm{\sut=16} and \mm{\sur=2}, relative to non-EVE (\mm{\sut=0}) condition.}

%
%
%

\noindent
The predictive correlation for the electroviscous correction factor (\mm{Y}, \fig\ref{fig:11}) in the asymmetrically charged contraction-expansion microfluidic device with dimensionless governing parameters (\mm{K,\sut,\sur, d_\text{c}=0.25}) is expressed as follows.
\begin{gather}
	Y = \alpha_4 Y_\text{ref}  \label{eq:Y}
	\\
	\alpha_4 = \beta_1 +  (\beta_2 + \beta_4 \sur)\sur + (\beta_3 + \beta_5 \gamma)\gamma + (\beta_6 +\beta_7 \sur )\sur \gamma
	\nonumber
	\\
	\beta_{\text{i}} = \sum_{{j}=1}^5 M_{\text{ij}} X^{({j}-1)};\quad X = K^{-1}; \quad  \gamma = \sut^{-1}; \quad 1\le i\le 7 \nonumber
\\
	M = \begin{bmatrix}
	1.1261 & -3.5179 & 15.482 & -29.732 & 21.028 \\
	-0.102 & 2.6144 & -6.5133 & 1.8619 & 6.3789 \\
	-0.8446 & 23.431 & -123.05 & 271.29 & -215.25 \\
	-0.0003 & 0.3004 & -5.8368 & 21.211 & -22.364 \\
	1.1785 & -29.947 & 157.07 & -338.59 & 259.39 \\
	0.4859 & -13.134 & 50.946 & -78.439 & 41.342 \\
	-0.024 & -0.5776 & 21.247 & -83.649 & 90.542 
	\end{bmatrix} \label{M:Y}
\end{gather}
where \mm{Y_\text{ref}} is the electroviscous correction factor \citep[\eqn (18) in ref.][]{dhakar2022electroviscous} for liquid flow through symmetrically charged (ref: \mm{\sur=1}) contraction-expansion slit.
The correlation coefficients (\mm{M_\text{ij}}, \eqn\ref{M:Y}) are statistically obtained by performing the non-linear regression analysis using the DataFit (version 9.0, free trial) for 135 data points, with 
(\mm{\delta\smin, \delta\smax, \delta\avg,R^2}) as (-2.27\%, 1.55\%, -0.24\%, 99.35\%) between  predicted (\eqn\ref{eq:Y}) and numerical (\fig\ref{fig:11}) values. 
%
%
\subsection{Pseudo-analytical model for pressure drop}
\noindent
The preceding discussion has shown the complex influences of the flow governing parameters on the hydrodynamic characteristics of the liquid electrolyte flow through asymmetrically charged non-uniform microfluidic device. However, predicting the hydrodynamic attributes using a simpler analytical model would be more convenient and robust to efficiently utilize the hydrodynamic characteristics in the engineering and design of relevant microfluidic systems.
\\\noindent
Previous studies have presented simple models to estimate the pressure drop (\mm{\Delta P}) in the symmetrically charged no-slip flow through contraction-expansion slit \citep{davidson2007electroviscous}, no-slip flow through contraction-expansion cylindrical pipe \citep{bharti2008steady}, and charge dependent slip flow through contraction-expansion slit \citep{dhakar2022electroviscous} microfluidic device.  
These studies \citep{dhakar2022electroviscous, davidson2007electroviscous, bharti2008steady} have estimated the total pressure drop (\mm{\Delta P}) in the steady incompressible laminar electrically neutral (\mm{\sut=0}) flow of Newtonian fluid through the contraction-expansion microfluidic slit device by summing up (a) the pressure drop in the independently uniform (i.e., \mm{\Delta P_\text{u}} in upstream, \mm{\Delta P_\text{c}} in contraction, and \mm{\Delta P_\text{d}} in downstream) rectangular sections by using the standard \textit{Hagen–Poiseuille} relation, and (b) the excess pressure drop (\mm{\Delta P_\text{e}}) due to sudden contraction/expansion accounted by the pressure drop through a thin (contraction ratio, \mm{d_\text{c} << 1}) orifice  \citep{Sisavath2002,davidson2007electroviscous,Pimenta2020}, as follows. 
\begin{gather}
	\Delta P_{\text{0,m}}=\left(\sum_{i=u,c,d}\Delta P_{\text{0,i}}\right) +\Delta P_{\text{0,e}}
	\label{eq:33}
\end{gather}
where, the subscript `0' indicates the electrically neutral (\mm{\sut=0}, or \mm{K=\infty}) flow, and the pressure drop terms are expressed as follows.
\begin{gather}
	\Delta P_{\text{0,i}} = \left(\frac{3}{Re}\right)\frac{{\Delta L_{\text{i}}}}{d_\text{i}^3};\qquad 
	\Delta P_{\text{0,e}} =\frac{16}{\pi d_{\text{c}}^2Re} \qquad \text{where}\qquad d_\text{i}=\frac{W_\text{i}}{W}
	\label{eq:33a,eq:34}
\end{gather}
%
%
Subsequently, the above model (\eqn\ref{eq:33}) is extended, and generalized pseudo-analytical model (\eqn\ref {eq:38}) is proposed in this study to obtain the pressure drop in the electroviscous (\mm{\sut>0}) flow through asymmetrically charged (\mm{\sur\neq 1}) microfluidic contraction-expansion (\mm{d_\text{c}=0.25}) slit device as follows.
\begin{gather}
	\Delta P_{\text{m}}= \Gamma_\text{evac} \Delta P_{0,\text{m}}
	=
	\Gamma_\text{evac}\left[\frac{3}{Re} \left(L_{\text{u}} +  \frac{L_{\text{c}}}{d_{\text{c}}^3} + L_{\text{d}} + \frac{16}{3\pi d_{\text{c}}^2} \right)\right]
	\label{eq:38}
\end{gather}
where, subscript `evac' denotes for the electroviscous and asymmetric charge (EVAC) effects, \mm{W} is used as the characteristic length in defining \mm{Re} (\eqn \ref{eq:1}), and to scale the length (upstream \mm{L_\text{u}}, contraction \mm{L_\text{c}}, downstream \mm{L_\text{d}}) variables.

\noindent
The correction factor (\mm{\Gamma_\text{evac}}) applied to \mm{\Delta P_{0,\text{m}}} for accounting the electroviscous (\mm{\sut>0}) and asymmetric charge (\mm{\sur\neq 1}) effects in \eqn(\ref{eq:38}) is statistically correlated as follows. 
\begin{gather}
	\Gamma_\text{evac} = 
	\beta_1 + (\beta_2 +\beta_4 \sur)\sur + (\beta_3 + \beta_5 \gamma)\gamma  + (\beta_6 + \beta_7 \sur)\sur\gamma 
\\
\beta_{\text{i}} = \sum_{{j}=1}^5 M_{\text{ij}} X^{({j}-1)};\quad X = K^{-1}; \quad  \gamma = \sut^{-1}; \quad 1\le i\le 7 \nonumber
\\
	M = \begin{bmatrix}
		0.8619 & 3.6321 & -11.5 & 14.871 & -5.855 \\
		-0.0925 & 2.1353 & 0.4404 & -21.49 & 30.052 \\
		2.1099 & -63.16 & 310.48 & -676.73 & 541.45 \\
		-0.0058 & 0.5014 & -7.9975 & 27.861 & -28.827 \\
		-6.4939 & 198.24 & -1072.8 & 2502.3 & -2098.6 \\
		0.4499 & -11.195 & 21.568 &	22.204 & -61.615 \\
		0.0004 & -1.4886 & 31.138 & -114.41 & 120.57	
	\end{bmatrix} \label{M:gevac}
\end{gather}
The correlation coefficients (\mm{M_\text{ij}}, \eqn\ref{M:gevac}) are statistically obtained by performing the non-linear regression analysis using the DataFit (version 9.0, free trial) for 135 data points, with 
(\mm{\delta\smin, \delta\smax, \delta\avg,R^2}) as (-1.43\%, 1.74\%, -0.19\%, 99.81\%) between predicted (\eqn\ref{eq:38}) and numerical (\tab\ref{tab:4}) values. 
\\\noindent
Subsequently, the electroviscous correction factor (\mm{Y}) can then be evaluated by using the pseudo-analytical model (\eqns\ref{eq:33} and \ref{eq:38}) as follows.
\begin{gather}
	Y_{\text{m}}=\frac{\Delta P_{\text{m}}}{\Delta P_{0,\text{m}}}
	\label{eq:40}
\end{gather}
\begin{figure}[tb]
	\centering
	\subfigure[pressure drop] {\includegraphics[width=0.49\linewidth]{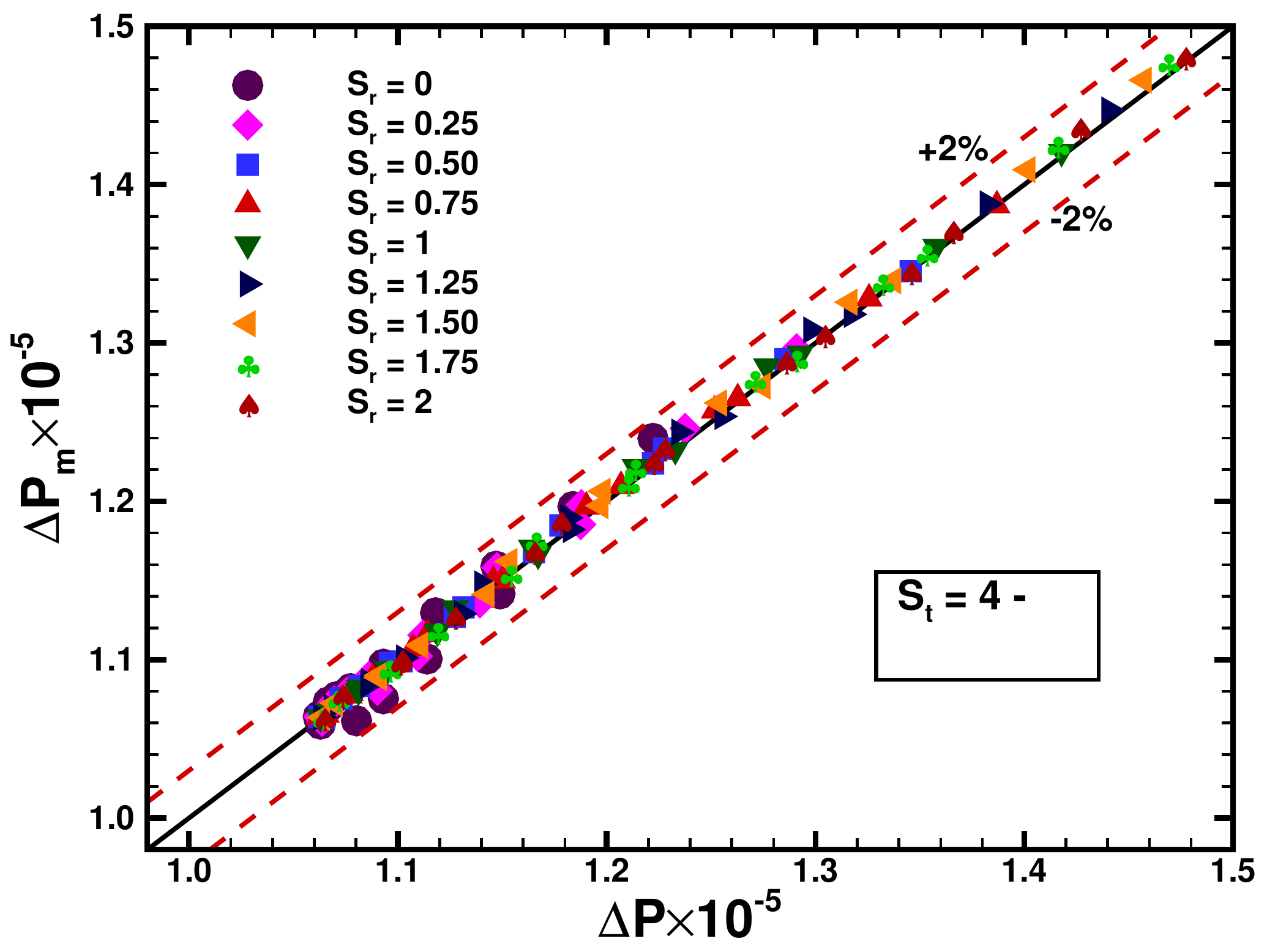}}
	\subfigure[electroviscous correction factor]
	{\includegraphics[width=0.49\linewidth]{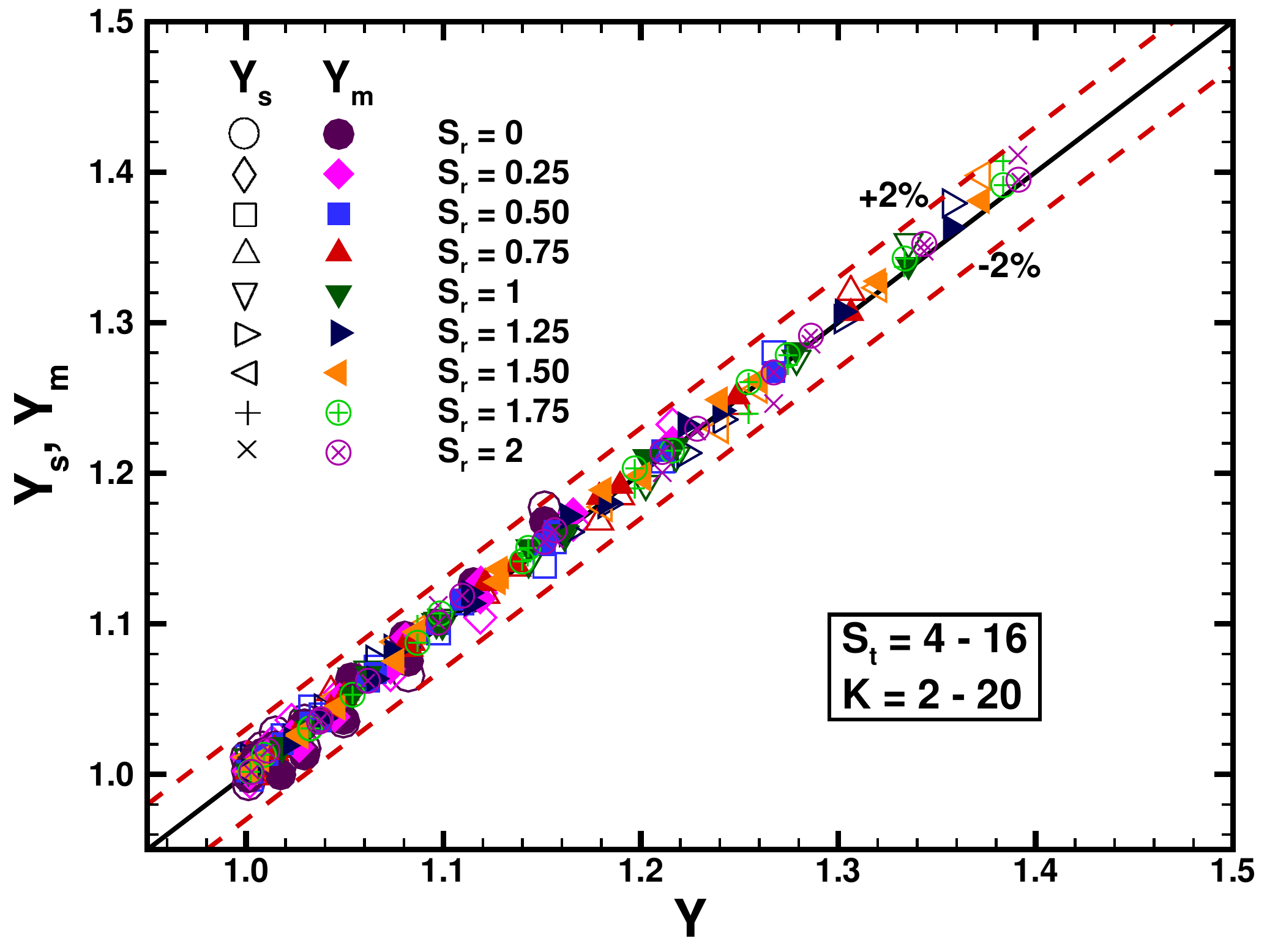}}
	\caption{Parity chart between the numerical and pseudo-analytical values of the (a) pressure drop, \mm{\Delta P} (\tab\ref{tab:4}) vs. \mm{\Delta P_{\text{m}}} (\eqn\ref{eq:38}), (b) electroviscous correction factor, \mm{Y} (\eqn\ref{eq:27}, \fig\ref{fig:11}) vs. \mm{Y_{\text{s}}} (\eqn\ref{eq:Y}) and \mm{Y_{\text{m}}} (\eqn\ref{eq:40}), for the considered parameters ($K$, $\sut$ and $\sur$).}
	\label{fig:12}
\end{figure} 

\noindent
\fig\ref{fig:12} represents the parity chart for pressure drop ($\Delta P$ vs $\Delta P_{\text{m}}$) and electroviscous correction factor ($Y$ vs $Y_{\text{m}}$ and $Y_{\text{s}}$) obtained numerically (\tab\ref{tab:4}, and \fig\ref{fig:11}) and analytically (\eqns\ref{eq:38} and \ref{eq:40}), for the broader ranges of flow governing parameters ($K$, $\sut$ and $\sur$) considered in the present work. 
A simpler model estimates both the pressure drop, and, hence the electroviscous correction factor $\pm 1-2\%$ within the numerical values. The difference between numerical and predicted results reduce with decreasing top wall surface charge density (\mm{\sut}), surface charge density ratio (\mm{\sur}), and thinning of EDL (increasing \mm{K}). Such a simple approach \citep{davidson2007electroviscous,bharti2008steady,bharti2009electroviscous,dhakar2022electroviscous} for the prediction of pressure drop, and thereby, electroviscous correction factor enables the use of present results in the design aspects for the microfluidic applications.
%
\section{Concluding remarks}
\noindent
In this study, electroviscous effects in the pressure-driven flow of electrolyte liquid through a asymmetrically charged contraction-expansion ($4$:$1$:$4$) slit microfluidic device have studied by using the Poisson's, Nernst-Planck and Navier-Stokes equations. These equations have been solved numerically using the finite element method (FEM) based COMSOL Multiphysics software to obtain the total potential ($U$), excess charge ($n^\ast$), induced electric field strength ($E_\text{x}$), velocity ($\myvec{V}$) and pressure ($P$) fields in the microfluidic device for broader ranges of parameters such as low Reynolds number ($Re=0.01$), Schmidt number ($\mathit{Sc}=1000$), inverse Debye length ($2\le K\le 20$), top wall surface charge density ($4\le \sut\le 16$) and surface charge ratio ($0\le \sur\le 2$). The main conclusions are obtained as follows: 

\begin{itemize}
	\item 
	\noindent
	The sudden contraction/expansion in the geometry changes excess charge, induced electric field strength, potential and pressure drop drastically in the contraction-section due to reduction in cross-section area. So, the potential and pressure gradients are maximum in the contraction-section.
	
	\item 
	\noindent
	With the change of symmetrically charged ($\sur=1$) condition to asymmetrically charged ($\sur\neq1$) condition, EDL overlaps {even at $K=8$ for high top wall surface charge density ($\sut=16$) and surface charge ratio ($\sur=1.50$)} in the microfluidic device flow.
	
	\item 
	\noindent
	The flow characteristics are depicted complex dependency on the governing parameters ($K$, $\sut$ and $\sur$). The total electrical potential and pressure drop change maximally by 197.45\% and 25.46\%, respectively, over the ranges of conditions. {The electroviscous correction factor (\mm{Y}) enhances maximally by 20.85\% (at \mm{K=2}, \mm{\sut=16}) and 34.16\% (at \mm{\sut=16}, \mm{\sur=2}) for the overall charge asymmetry (\mm{0\le \sur\le 2}), and \mm{K} variation from 20 to 2, respectively. Further, electroviscous correction factor maximally increases by 39.13\% at \mm{K=2}, \mm{\sut=16} and \mm{\sur=2}, relative to non-EVE (\mm{\sut=0}) condition.} Thus, asymmetry in the surface charge enhances the electroviscous effects in the microfluidic device flow.
	
	\item 
	\noindent
	A simpler analytic model is developed, based on the Poiseuille flow for each uniform section (upstream, contraction and downstream sections) pressure drop and excess pressure loss due to sudden contraction and expansion by creeping flow for thin orifice. This simpler model estimates the pressure drop 1-2\% within the numerical results value. 
	
	\item 
	\noindent 
	The numerical results are presented in terms of mathematical correlations for easy use and expending these results for broader ranges of parameters. Thus, the numerical results correlations and simplicity and robustness of simpler analytical model enable the uses of present numerical results in the practical microfluidic applications.
\end{itemize}
%
%
%
\section*{Declaration of Competing Interest}
\noindent 
The authors declare that they have no known competing financial interests or personal relationships that could have appeared to influence the work reported in this paper.
%
\section*{Acknowledgements}
\noindent 
R.P. Bharti would like to acknowledge Science and Engineering Research Board (SERB), Department of Science and Technology (DST), Government of India (GoI) for the providence of the MATRICS grant (File no. MTR/2019/001598). 
%
\begin{spacing}{1.1}
\input{Nomenclature.tex}

\printnomenclature
\end{spacing}
%
%
\appendix
\setcounter{figure}{0} 
\setcounter{table}{0} 
\noindent 
\bibliography{references}
%
%
%
%
%
%
%
%
%
%
%
%
\end{document}

%% file: nomenclature.tex
\fontsize{10}{10pt}\selectfont
 \nomenclature[g0]{\textit{Greek letters}}{}
 \nomenclature[d0]{\textit{Dimensionless groups}}{}
 \nomenclature[s0]{\textit{Subscripts and Superscripts}}{}
 \nomenclature[z0]{\textit{Abbreviations}}{}
%
\nomenclature[zcfd]{CFD}{computational fluid dynamics}
\nomenclature[zedl]{EDL}{electrical double layer}
\nomenclature[zeve]{EVE}{electroviscous effect}
\nomenclature[zevf]{EVF}{electroviscous flow}
\nomenclature[zfem]{FEM}{finite element method}
\nomenclature[zfvm]{FVM}{finite volume method}
\nomenclature[zpdes]{PDEs}{partial differential equations}
\nomenclature[zpdps]{PDF}{pressure-driven flow}
\nomenclature[zsaes]{SAEs}{simultaneous algebraic equations}
%
\nomenclature[adc]{$d_{\text{c}}$}{contraction ratio ($=W_{\text{c}}/W$), --}
\nomenclature[aDj]{$\mathcal{D}_{j}$}{diffusivity of the ions of type j, assumed equal ($\mathcal{D}_{+}=\mathcal{D}_{-}=\mathcal{D}$), m$^2$/s}
\nomenclature[ae]{$e$}{elementary charge of a proton ($=1.602176634\times 10^{-19}$), C or A.s}
\nomenclature[aE]{$E_{\text{x}}$}{induced electric field strength ($=-\partial U/\partial x$), V/m or --}
\nomenclature[afj]{$\mathbf{f_\text{j}}$}{flux density of the ions of type j (\eqn\ref{eq:9}), 1/(m$^2$.s)}
\nomenclature[aIc]{$I_{\text{c}}$}{conduction current density (\eqn\ref{eq:7}), A/m$^2$ or --}
\nomenclature[aId]{$I_{\text{d}}$}{diffusion current density (\eqn\ref{eq:7}), A/m$^2$ or --}
\nomenclature[aIs]{$I_{\text{s}}$}{streaming current density (\eqn\ref{eq:7}), A/m$^2$ or --}
\nomenclature[akB]{$k_{\text{B}}$}{Boltzmann constant ($=1.380649\times 10^{-23}$), J/K}
\nomenclature[aLc]{$L_{\text{c}}$}{length of contraction section, m or --}
\nomenclature[aLd]{$L_{\text{d}}$}{length of downstream outlet section, m or --}
\nomenclature[aLu]{$L_{\text{u}}$}{length of upstream inlet section, m or --}
\nomenclature[an+]{$n_{+}$}{local number density of positive ions (\eqn\ref{eq:6}), 1/m$^3$ or --}
\nomenclature[an-]{$n_{-}$}{local number density of negative ions (\eqn\ref{eq:6}), 1/m$^3$ or --}
\nomenclature[an0]{$n_{0}$}{bulk density of the ions of type j, 1/m$^3$}
\nomenclature[anj]{$n_{j}$}{local number density of the ions of type j, 1/m$^3$}
\nomenclature[ans]{$n^*$}{excess charge ($=n_{+}-n_{-}$), 1/m$^3$ or --}
\nomenclature[ant]{$n^{**}$}{normalized excess charge,  --}
\nomenclature[aP]{$P$}{pressure, Pa or --}
\nomenclature[aT]{$T$}{temperature, K}
\nomenclature[aU]{$U$}{total electrical potential, V or --}
\nomenclature[aU]{$U_{\text{c}}$}{characteristic total electrical potential, V}
\nomenclature[aV]{$\mathbf{V}$}{velocity vector, m/s or --}
\nomenclature[aVa]{$\overline{V}$}{average velocity of the fluid at the inlet, m/s}
\nomenclature[aVx]{$V_x$}{x-component of the velocity, m/s or --}
\nomenclature[aVy]{$V_y$}{y-component of the velocity, m/s or --}
\nomenclature[aW]{$W$}{cross-sectional width of inlet and outlet sections, m}
\nomenclature[aWc]{$W_{\text{c}}$}{cross-sectional width of contraction section, m}
\nomenclature[ax]{$x$}{streamwise coordinate, --}
\nomenclature[ay]{$y$}{transverse coordinate, --}
\nomenclature[aY]{$Y$}{electroviscous correction factor (\eqns\ref{eq:27}, and \ref{eq:40}), --}
\nomenclature[azj]{$z_{j}$}{valency of the ions of type j, assumed equal ($z_{+}=\rev{-}z_{-}=z$), --}
%
%
\nomenclature[gdP]{$\Delta P$}{pressure drop (\eqns\ref{eq:38}), --}
\nomenclature[geps0]{$\varepsilon_{\text{0}}$}{permittivity of free space (i.e. vaccum), F/m or C/(V.m)}
\nomenclature[gepsr]{$\varepsilon_{\text{r}}$}{dielectric constant (absolute or relative permittivity) of the electrolyte liquid, --}
\nomenclature[glambdad]{$\lambda_{\text{D}}$}{Debye length ($=K^{-1}$), --}
\nomenclature[gmu]{$\mu$}{viscosity, Pa.s}
\nomenclature[gmueff]{$\mu_\text{eff}$}{effective or apparent viscosity, Pa.s}
\nomenclature[gpsi]{$\psi$}{EDL potential (\eqn\ref{eq:2a}), V or --}
\nomenclature[grho]{$\rho$}{density of fluid, kg/m$^3$}
\nomenclature[grhoe]{$\rho_{\text{e}}$}{charge density of liquid, C/m$^3$}
\nomenclature[gsigmab]{$\sigma_\text{b}$, $\sigma_\text{t}$}{surface charge density, C/m$^2$}
%
%
\nomenclature[dbeta]{$\mathit{\beta}$}{liquid parameter (\eqn\ref{eq:1}), --}
\nomenclature[dK]{$\mathit{K}$}{inverse Debye length (\eqn\ref{eq:1}), --}
\nomenclature[dPe]{$Pe$}{Peclet number ($={Re}\times\mathit{Sc}$) (\eqn\ref{eq:1}), --}
\nomenclature[dRe]{$Re$}{Reynolds number (\eqn\ref{eq:1}), --}
\nomenclature[dSa]{$\mathit{S_\text{b}}$, $\mathit{S_\text{t}}$}{surface charge density (\eqn\ref{eq:12}), --}
\nomenclature[dSc]{$\mathit{Sc}$}{Schmidt number (\eqn\ref{eq:1}), --}
\nomenclature[dSb]{$\mathit{S_\text{r}}$}{surface charge density ratio (\eqn\ref{eq:12}), --}
%
\nomenclature[sz]{$0$}{without electroviscous effects}
\nomenclature[sc]{$c$}{contraction}
\nomenclature[sd]{$d$}{downstream}
\nomenclature[se]{$e$}{extra or excess}
\nomenclature[sf]{$evac$}{electroviscous and asymmetric charge}
\nomenclature[sm]{$m$}{mathematical}
\nomenclature[ss]{$s$}{statistical}
\nomenclature[su]{$u$}{upstream}
%